\documentclass[12pt,a4paper,twoside,fleqn]{article} 

\usepackage{latexsym} 
\usepackage{amsfonts} 

\begin{document} 

\newcommand{\sumint}[1]
{\begin{array}{c}
\\
{{\textstyle\sum}\hspace{-1.1em}{\displaystyle\int}}\\
{\scriptstyle{#1}}
\end{array}} 

\newcommand{\ket}[1]{\left|#1\right>} 
\newcommand{\bra}[1]{\left<#1\right|} 
\newcommand{\braket}[2]
{\left<#1|#2\right>} 

\newcommand{\ner}[1]{\left|#1\right)} 
\newcommand{\inn}[1]{\left(#1\right|} 
\newcommand{\inner}[2]
{\left(#1|#2\right)}

\newcommand{\lar}[1]{\left|#1\right\}} 
\newcommand{\sca}[1]{\left\{#1\right|} 
\newcommand{\scalar}[2]
{\left\{#1|#2\right\}}  

\newcommand{\f}[1]{\mbox{\boldmath$#1$}}

\title{On the Particle Definition in the presence of Black Holes}

\author{Ralf Sch\"utzhold\\
Institut f\"ur Theoretische Physik, Technische Universit\"at Dresden\\
D-01062 Dresden, Germany\\
Electronic address : {\tt schuetz@theory.phy.tu-dresden.de}}

\date{\today}

\maketitle
 
\begin{abstract} 
 
A canonical particle definition via the diagonalisation of the Hamiltonian 
for a quantum field theory in specific curved space-times is presented.
Within the provided approach radial ingoing or outgoing Minkowski particles 
do not exist. 
An application of this formalism to the Rindler metric recovers the 
well-known Unruh effect. 
For the situation of a black hole the Hamiltonian splits up into two
independent parts accounting for the interior and the exterior domain,
respectively. 
It turns out that a reasonable particle definition may be accomplished
for the outside region only. 
The Hamiltonian of the field inside the black hole is unbounded from
above and below and hence possesses no ground state.
The corresponding equation of motion displays a linear
global instability.
Possible consequences of this instability are discussed and its
relations to the sonic analogues of black holes are addressed.     

\end{abstract} 
 
\bigskip 
 
PACS-numbers:  
04.70.Dy, 
04.62.+v, 
10.10.Ef, 
03.65.Db  

\vspace{0.1cm}

\newpage
  
\section{Introduction}\label{intro}

At present we know two very fundamental and effectual theories in order to 
describe nature, quantum theory and general relativity.
But a satisfactory unification of these distinct theories is still missing.
One possibility to achieve some progress towards this aim is expected to be
provided by the consideration of quantised fields in given classical 
space-times. 
Within this treatment the metric plays the role of an external background 
field.
Various investigations have been devoted to this topic during the last 
decades, to mention only some of the most important initial papers in
chronological order:
In 1972 Fulling \cite{fulling} noticed the non-uniqueness of the
particle interpretation in curved space-times 
(which may be regarded as the basis for several effects).
Two years later Hawking \cite{hawking} found out that black holes are not 
completely black but possess a thermal behaviour caused by the conversion of 
the initial vacuum fluctuations.   
A short time after this striking discovery  Davies \cite{davies} showed that 
also a uniformly accelerated mirror -- treated as a mirror at rest in the 
Rindler metric -- creates a thermal spectrum.
In 1976 Unruh \cite{unruh} recognised the fact that even a uniformly 
accelerated observer in the Minkowski vacuum feels environed by a thermal 
bath.
Many examinations have been accomplished since these basic papers, see e.g.  
\cite{boulware}--\cite{wipf} and references therein.
Now black holes are very interesting touchstones in order to test
candidates for the theory that unifies general relativistic and
quantum aspects. The representations of black holes within the
underlying theory are expected to reproduce their main properties. 
  
The main intention of this article is to provide a canonical approach
to quantum field theory in specific curved space-times and a related
particle definition, together with an investigation of the consequences of 
this formalism.

This paper is organised as follows:
In Section \ref{eom} we set up the equations which describe the quantum field 
under consideration together with the assumptions necessary for the particle 
definition.   
Based on methods of functional analysis we propose in Sec. \ref{pdef}
a canonical approach to the particle definition via the diagonalisation of the 
Hamiltonian.
This procedure is applied to some flat space-times in Section \ref{flat} in 
order to elucidate the underlying mechanism.
Sec. \ref{schwarz} considers the space-time of a black hole and
Section \ref{ext} presents the particle definition for the exterior domain.   
In Sec. \ref{int} we examine the quantised field inside
the black hole and deduce the unstable behaviour of its time-evolution.
Possible consequences of this instability are discussed in Section
\ref{back} and its implications for the sonic analogues of black holes
are pointed out in Section \ref{sonic}. 
We shall close with a summary, some discussions and an outline.

Throughout this article natural units with $G=\hbar=c=k_{\rm B}=1$ 
will be used.
Lowercase Greek indices such as $\mu,\nu$ vary from 0 (time) to
3 (space) and describe space-time components.
Lowercase Roman indices $i,j$ range from 1 to 3 and 
label only the space. 
(For both we employ the Einstein sum convention.)
Uppercase Roman indices $I,J\in\mathbb N$ may assume all
natural numbers while uppercase Greek indices like
$\Gamma,\Lambda$ may be more general, e.g. continuous.

\section{Equations of motion}\label{eom}

We consider a minimally coupled, massless and neutral (i.e. real) scalar 
field $\Phi$ whose propagation in the space-time $(M,g^{\mu\nu})$ is 
described by the action 
\begin{eqnarray}
\label{action}
{\cal A}=\int\limits_M d^4x\,\frac{\sqrt{-g}}{2}\,
g^{\mu\nu}(\partial_\mu\Phi)(\partial_\nu\Phi)
\,,
\end{eqnarray}
with $g=\det(g_{\mu\nu})$.
Possible potential terms like a mass term $m^2\,\Phi^2$ or a conformal
coupling term ${\cal R}\,\Phi^2/6$ 
(where ${\cal R}={\cal R}^\mu_\mu$ denotes the Ricci scalar)
do not alter the main conclusions, see Secs. \ref{Kop} and \ref{int}.  
The same holds true for a charged and thus complex field $\Phi$ and $\Phi^*$. 
For reasons of simplicity and considering the scalar field $\Phi$ as a 
model for the photon field we restrict ourselves to the most simple
action in Eq. (\ref{action}).
 
Provided that the spatial surface terms arising from the integration by parts 
vanish the variation of the action $\delta{\cal A}=0$ leads to the 
Klein-Fock-Gordon equation
\begin{eqnarray}
\Box\Phi=\frac{1}{\sqrt{-g}}\,
\partial_\mu\left(\sqrt{-g}\,g^{\mu\nu}\partial_\nu\Phi\right)=0
\,.
\end{eqnarray}
The corresponding inner product is defined by
\begin{eqnarray}
\inner{\psi}{\phi}\stackrel{\rm def}{=} i\int\limits_\Sigma d\Sigma^\mu  
\;\psi^*\stackrel{\leftrightarrow}{\partial}_\mu\phi
\end{eqnarray}
with
$\psi\stackrel{\leftrightarrow}{\partial}_{\mu}\,  
\phi=\psi\,\partial_{\mu}\,\phi-\phi\,\partial_{\mu}\,\psi$.
In this definition the surface element $d\Sigma_\mu$ already contains
factors like $\sqrt{-g_\Sigma}$ with the result
$d\Sigma_\mu dx^\mu=d^4x\,\sqrt{-g}$.
For functions which fulfil the Klein-Fock-Gordon equation 
$\Box\psi=\Box\phi=0$ the inner product is independent of the 
special surface $\Sigma$, cf. for instance \cite{ford}.
To see that, one has to use Gauss' law
\begin{eqnarray}
\oint\limits_{\partial M_\Sigma} d\Sigma_{\mu}\;A^\mu=
\int\limits_{M_\Sigma} d^4x\;\sqrt{-g}\;\nabla_\mu\,A^\mu  
\end{eqnarray}
and again to require vanishing spatial surface terms.

\subsection{Preconditions}\label{precond}

Now we have to specify the assumptions which are necessary for an appropriate 
particle definition. At first we demand a {\em strongly causal space-time $M$}.
This condition ensures the essential physical principle of distinguishing 
cause and effect and forbids (for instance) the occurrence of closed time-like
curves, cf. \cite{ellis}.

As another requirement we impose a {\em static metric} of the space-time $M$ 
\begin{eqnarray}
ds^2=g_{\mu\nu}dx^\mu dx^\nu=
g_{00}(\f{r})dt^2+g_{ij}(\f{r})dx^i dx^j
\,.
\end{eqnarray}
These two assumptions allow to factorise the space-time 
$M={\mathbb R}\otimes G$ into time $t\in\mathbb R$ and space $\f{r}\in G$
with an open domain $G\subseteq{\mathbb R}^3$.
The Killing vector corresponding to the time translation symmetry permits 
the definition of a conserved energy. 
This fact is substantial for a physical reasonable particle
definition, see also Sec. \ref{energy} below.
On the other hand, the selection of a particular Killing vector
refers to a class of associated observers whose time evolution 
is generated by this vector field. 
In general different Killing vectors generate distinct particle
definitions applying for the different observers, 
see Sec. \ref{unruh} below.

The third precondition we need is called {\em non-degenerated signature}.
This fixes the signature of the metric inside the domain $G$
\begin{eqnarray}
\forall\,\f{r}\in G\quad : \quad g_{00}(\f{r})>0
\quad ; \quad \left(g_{ij}(\f{r})\right)_{ij}<0
\,.
\end{eqnarray}
In the latter inequality $\left(g_{ij}(\f{r})\right)_{ij}$
has to be understood as a matrix (and not as the single components), i.e.
$\forall\f{p}\in{\mathbb R}^3:\f{p}^2>0\rightarrow p_ig^{ij}p_j<0$.
Both quantities ($g_{00}$ and $g_{ij}$) are continuous and regular 
inside $G$ but may diverge or vanish by approaching the boundary $\partial G$.
This might be the case for a horizon situated at $\partial G$.
(A $g^{00}$-component of the metric that vanishes over a finite
volume would create primary constraints, see e.g. \cite{teitelboim}.)

As demonstrated above the possibility of performing the integration by parts 
is a really important issue. 
Accordingly, our last assumption is a {\em physical complete region $G$}. 
This simply enforces the vanishing of the surface terms.
The occurrence of such boundary contributions always indicates the 
interaction with a system behind the surface.
Such a region would not be physical complete. 
It should be noted here that the validity of the integration by parts also
includes periodicity in angular coordinates, such as
$f(\varphi)=f(\varphi+2\pi)$.
For the above specified space-time $M={\mathbb R}\otimes G$ the spatial 
surface terms read  
\begin{eqnarray}
\int\limits_{\partial G} dS_i\,\Phi\,g^{ij}\,\partial_j\Phi=0
\,.
\end{eqnarray}
There are several ways to achieve the equation above.
For Dirichlet boundary conditions one would demand $\Phi=0$ at $\partial G$ 
and for Neumann type $dS_i\,g^{ij}\,\partial_j\Phi=0$ at $\partial G$.
But there is also a third possibility for the disappearance of the surface 
terms, namely if the components of the metric themselves which are orthogonal 
to the surface $dS_i$ approach zero at $\partial G$, i.e. 
$dS_i\,g^{ij}(\partial G)=0$.
As stated above this might be the case for a horizon situated at the boundary 
$\partial G$.

Strictly speaking, there exist various definitions of a horizon, such
as the event, apparent, Cauchy, particle, and putative horizon, 
cf. \cite{ellis} and \cite{visser}.
The definition of the particle horizon refers to a special observer
at a given time-like word-line whereas the other horizons can be
defined in an observer-independent way. 
Hence, the vanishing of the spatial surface terms (without constraints
on the fields) implies a priori only a particle horizon at $\partial G$.
However, with some additional requirements on the space-time,
for instance spherical symmetry and asymptotic flatness, the (particle)
horizon at $\partial G$ meets the other definitions as well.

After having established the properties of the space-time $M$,
we arrive at the conclusion that it is globally hyperbolic 
(i.e. strongly causal and complete, cf. \cite{ellis})
and the spatial domain $G$ represents for every time $t$ a Cauchy surface.

\subsection{Energy}\label{energy}

For a time-independent metric the Noether theorem demands the 
existence of a conserved energy. 
The energy-momentum tensor for the scalar field reads
\begin{eqnarray}
\label{Tmunu}
T_{\mu\nu}=\frac{2}{\sqrt{-g}}\,\frac{\delta\cal A}{\delta g^{\mu\nu}}=
(\partial_\mu\Phi)(\partial_\nu\Phi)-\frac{g_{\mu\nu}}{2}\,
(\partial_\gamma\Phi)(\partial^\gamma\Phi)
\,.
\end{eqnarray}
By virtue of the Klein-Fock-Gordon equation $\Box\Phi=0$ the covariant 
divergence of the energy-momentum tensor vanishes
\begin{eqnarray}
\nabla_\mu\,T^\mu_\nu=\frac{1}{\sqrt{-g}}\;
\partial_\mu\left(\sqrt{-g}\;T^\mu_\nu\right)-
\frac{1}{2}\;T^{\alpha\beta}\,\partial_\nu\,g_{\alpha\beta}=0
\,.
\end{eqnarray}
In general this covariant equation does not lead to any conserved quantities 
due to the exchange of energy and momentum between the gravitational and 
the scalar field (second term).
But for a stationary metric ($\partial_0\,g_{\alpha\beta}=0$)
it is possible to construct a conserved energy flux $j^\mu$ utilising the 
($\nu=0$)-components
\begin{eqnarray}
\partial_\mu\,j^\mu=
\partial_\mu\left(\sqrt{-g}\;T^\mu_0\right)=0
\,.
\end{eqnarray}
This local conservation law allows for the introduction of a conserved
energy as a global quantity via
\begin{eqnarray}
E\stackrel{\rm def}{=}\int\limits_G d^3x\;\sqrt{-g}\;T^0_0
=\int\limits_G d^3\f{r}\;T^0_0
\,.
\end{eqnarray}
For a Minkowski space-time where $T^{00}_{ }=T^{0}_{0}=T_{00}$ holds this 
definition coincides (of course) with the usual energy.
Another argument for the above defined energy for being the correct choice
is the following:
Starting from the action ${\cal A}$ we may define the Lagrange function $L$ 
such that
\begin{eqnarray}
{\cal A}\stackrel{\rm def}{=}\int dt\;L
\end{eqnarray}
holds.
The Hamilton function $H$ as the Legendre transform of this Lagrange function 
exactly coincides with the energy of the field $H=E$.

\section{Particle definition}\label{pdef}

To provide a canonical definition of particles one has to indicate which 
properties the particles should exhibit.
For a free (linear) field we expect the particles to evolve independently 
and to carry a certain energy.
As shown in Section \ref{energy}, for a static metric the energy $E$ 
of the field $\Phi$ and its Hamilton function $H$ coincide.
Consequently, both requirements can be satisfied by the diagonalisation 
of $E=H$ or, equivalently, the Lagrange function $L$.  
Having defined the particles via diagonalisation of $H=E$, the corresponding 
vacuum $\ket{0}$ coincides with the ground state of the Hamiltonian and 
the energy. 
Of course, the procedure described above does not represent the only
one possibility to accomplish the particle definition.
Another approach is based on the "one-particle structure" of classical 
solutions of the field equation, see e.g. 
\cite{wald-p,reed+simon,ashtekar,kay,dimock}
and the remarks in Sec. \ref{vacuum}.

According to the definition in Section \ref{energy} the Lagrange function
governing the dynamics of the field reads
\begin{eqnarray}
\label{lagGij}
L=
\frac{1}{2}\int\limits_G d^3\f{r}\,g^{00}(\f{r})\,\dot\Phi^2+
\frac{1}{2}\int\limits_G d^3\f{r}\,g^{ij}(\f{r})
(\partial_i\Phi)(\partial_j\Phi)
\,,
\end{eqnarray}
where $d^3\f{r}$ denotes the spatial integration with the volume element
$d^3\f{r}=\sqrt{-g}\,d^3x$. 

To diagonalise this expression one has to deal with an elliptic partial 
differential operator which requires some functional analysis.
All of the used theorems can be found in \cite{reed+simon,weidmann}
and are not cited explicitly in the following.

\subsection{Hilbert space theory}\label{hilbert}

To work with mathematically well-defined quantities we have to set up some 
definitions.
$C^\infty_0(G)$ denotes the set of all infinitely differentiable functions
$u\,:\,G\rightarrow\mathbb R$ of compact support inside the open domain $G$.   
For two functions of this kind $u,v\in C^\infty_0(G)$ we define a scalar 
product via
\begin{eqnarray}
\label{scalarproduct}
\scalar{u}{v}_1
\stackrel{\rm def}{=}\int\limits_G d^3\f{r}\,g^{00}(\f{r})\,u^*(\f{r})\,v(\f{r})
\,.
\end{eqnarray}
The assumption of a non-degenerated signature in Section \ref{precond}
is essential for this definition.
Without a positive $g^{00}$ the above expression would be a pseudo-scalar 
product instead of a scalar product with $\scalar{u}{u}=0\leftrightarrow u=0$.
The latter property is necessary for investigations concerning convergence.    
As every scalar product induces a norm $||u||^2=\scalar{u}{u}$ it is now
possible to define a Hilbert space as the completion of all $C^\infty_0(G)$ 
functions with respect to this norm  
\begin{eqnarray}
L_2(G,g^{00})\stackrel{\rm def}{=}
\overline{C^\infty_0(G)}^{\,\scalar{\cdot}{\cdot}_1}
\,.
\end{eqnarray}
Because every $C^\infty_0(G)$-function can be $L_2(G,g^{00})$-approximated 
by linear combinations of step functions, this Hilbert space is separable.

The same procedure may be performed for vector-valued functions
$\f{u}\,:\,G\rightarrow{\mathbb R}^3$.
Again we may define a scalar product for two smooth functions of compact 
support $\f{u},\f{v}\in[C^\infty_0(G)]^3$ due to the non-degenerated signature 
\begin{eqnarray}
\scalar{\f{u}}{\f{v}}_3
\stackrel{\rm def}{=}
-\int\limits_G d^3\f{r}\,g^{ij}(\f{r})\,u_i^*(\f{r})\,v_j(\f{r})\,,
\end{eqnarray}
and in analogy the corresponding Hilbert space reads
\begin{eqnarray}
L_2^3(G,g^{ij})\stackrel{\rm def}{=}\overline{[C^\infty_0(G)]^3}
^{\,\scalar{\cdot}{\cdot}_3}
\,.
\end{eqnarray}
The advantage of the scalar products defined in such a way becomes evident if 
we use the linear partial differential operator
\begin{eqnarray}
{\cal D}:\,C^\infty_0(G)\subset L_2(G,g^{00}) 
& \rightarrow & L_2^3(G,g^{ij})
\nonumber\\
\phi(\f{r}) & \rightarrow & \left(\partial_i\phi(\f{r})\right)_i
\end{eqnarray}
to cast the Lagrange function into the simple form
\begin{eqnarray}
L=\frac{1}{2}\scalar{\dot\Phi}{\dot\Phi}_1
-\frac{1}{2}\scalar{{\cal D}\Phi}{{\cal D}\Phi}_3
\,.
\end{eqnarray}
Nevertheless, this is still not a representation which is suitable for 
diagonalisation.  
For that purpose we have to perform the spatial integration by parts 
(see Section \ref{eom}). 
In terms of functional analysis this means the construction of the adjoint 
operator.
The domain of definition ${\rm Def}({\cal D})=C^\infty_0(G)$ of the 
${\cal D}$-operator is dense in $L_2(G,g^{00})$.
As a consequence, its adjoint ${\cal D}^\dagger $ exists as a linear operator
${\cal D}^\dagger \,:\,{\rm Def}({\cal D}^\dagger )\subset
L_2^3(G,g^{ij})\rightarrow L_2(G,g^{00})$. 
For $[C^\infty_0(G)]^3$-functions the spatial integration by parts is always 
possible.
Accordingly, the domain of definition of the adjoint ${\cal D}^\dagger $
contains these functions 
$[C^\infty_0(G)]^3\subset{\rm Def}({\cal D}^\dagger )$
and is thereby also dense in $L_2^3(G,g^{ij})$.
Therefore the twice adjoint ${\cal D}^{\dagger \dagger }$ exists as a linear 
operator as well
${\cal D}^{\dagger \dagger }\,:\,{\rm Def}({\cal D}^{\dagger \dagger })
\subset L_2(G,g^{00})\rightarrow L_2^3(G,g^{ij})$. 
Of course, these operators describe physical reality only if one ensures the 
possibility of the spatial integration by parts via physical reasons as done 
in Section \ref{precond}. 

\subsection{${\cal K}$-operator}\label{Kop}

Now we are in the position to cast the Lagrange function into a form which
can be utilised for the diagonalisation of the system.
With the definition of the elliptic partial differential operator  
${\cal K}\stackrel{\rm def}{=}{\cal D}^\dagger {\cal D}^{\dagger \dagger }$
(see also \cite{fulling,fulling-buch}) we arrive at
\begin{eqnarray}
\label{lagK}
L
&=&
\frac{1}{2}\scalar{\dot\Phi}{\dot\Phi}_1
-\frac{1}{2}\sca{\Phi}{\cal D}^\dagger {\cal D}^{\dagger \dagger }
\lar{\Phi}_1
\nonumber\\
&\stackrel{\rm def}{=}&
\frac{1}{2}\scalar{\dot\Phi}{\dot\Phi}_1
-\frac{1}{2}\sca{\Phi}{\cal K}\lar{\Phi}_1
\,.
\end{eqnarray}
Every linear operator of the form 
${\cal K}={\cal D}^\dagger {\cal D}^{\dagger \dagger }$
is non-negative and self-adjoint and thus can be diagonalised.
Let us study the domain of definition ${\rm Def}({\cal K})$ of this operator.
The twice adjoint operator ${\cal D}^{\dagger \dagger }$ is the closure of 
the original operator ${\cal D}$, i.e. 
$\overline{\cal D}={\cal D}^{\dagger \dagger }$. 
Its domain of definition is the completion of all $C^\infty_0(G)$-functions
\begin{eqnarray}
{\rm Def}(\overline{\cal D})=
\overline{C^\infty_0(G)}^{\,\scalar{\cdot}{\cdot}_{\cal D}}
\end{eqnarray}
with respect to the graph scalar product which is defined via
\begin{eqnarray}
\scalar{u}{v}_{\cal D}\stackrel{\rm def}{=}
\scalar{u}{v}_1+
\scalar{{\cal D}u}{{\cal D}v}_3
\,.
\end{eqnarray}
One observes that the operator ${\bf 1}+{\cal K}$ is exactly the Friedrich 
extension (which is self-adjoint, see \cite{weidmann}) of the original 
operator ${\bf 1}+{\cal K}\mid_{C^\infty_0(G)}$ mediated via the graph scalar 
product. 
As a result, if the domain $G$ has boundaries $\partial G$ with Dirichlet 
boundary conditions, these boundary conditions are already incorporated into 
the domain of definition of the operators $\overline{\cal D}$ and
${\cal K}$, i.e.
\begin{eqnarray}
\phi\in{\rm Def}({\cal K})\subset{\rm Def}(\overline{\cal D})
\rightarrow\phi(\partial G)=0
\,.
\end{eqnarray}
To incorporate Neumann boundary conditions one has to start with an
operator like $(u^i)^i\rightarrow\partial_iu^i$ and to proceed in
the same way.

As mentioned in Sec. \ref{eom}, additional potential terms do not
alter the main conclusions. If we assume the scalar curvature to be
a bounded $-m^2 \leq {\cal R} \leq {\cal R}_{\rm Max}$ and smooth 
${\cal R} \in C^\infty(G)$ function we may introduce a new operator via
\begin{eqnarray}
{\cal B}:\,L_2(G,g^{00}) 
& \rightarrow & L_2(G,g^{00})
\nonumber\\
\phi(\f{r}) & \rightarrow & \left(m^2+{\cal R}\right)\phi(\f{r})
\,.
\end{eqnarray}
Obviously this operator is bounded, non-negative, and self-adjoint.
In addition, since 
${\rm Def}({\cal D}^\dagger\overline{\cal D})
\subset{\rm Def}({\cal B})=
L_2(G,g^{00})$, 
we may define a modified $\cal K$-operator via
\begin{eqnarray}
{\cal K}\stackrel{\rm def}{=}{\cal D}^\dagger\overline{\cal D}
+{\cal B}\mid_{{\rm Def}({\cal D}^\dagger\overline{\cal D})}
\,,
\end{eqnarray}
which is still self-adjoint and non-negative.
 
\subsection{Spectral theory}\label{stheory}

As mentioned above, every self-adjoint operator can be diagonalised.
One way to reveal this statement in a more explicit form is the following 
theorem:
For every self-adjoint operator $\cal K$ there exists a spectral family
$\cal E$ of orthogonal projections with
\begin{eqnarray}
{\cal K}=\int\lambda\;d{\cal E}(\lambda)
\,.
\end{eqnarray}
$d{\cal E}(\lambda)$ contributes only for values $\lambda$ being in the 
spectrum $\sigma({\cal K})$ of the $\cal K$-operator 
$\lambda\in\sigma({\cal K})$.
The spectrum $\sigma({\cal K})$ of an operator $\cal K$ contains all complex 
numbers $z\in\mathbb C$ for which the resolvent 
${\cal R}(z)\stackrel{\rm def}{=}(z-{\cal K})^{-1}$ does not exist, 
i.e. $(z-{\cal K})^{-1}$ is 
not a well and densely defined and bounded operator.
For a self-adjoint and non-negative operator ${\cal K}$ the spectrum is 
purely real and non-negative $\sigma({\cal K})\subseteq{\mathbb R}_+$.
It splits up into two parts, the point spectrum $\sigma_{\rm p}$ and the
continuous spectrum $\sigma_{\rm c}$.
The point spectrum is the set of all proper eigenvalues $\lambda$
corresponding to proper eigenfunctions
\begin{eqnarray}
\sigma_{\rm p}=\left\{
\lambda\in{\mathbb C}\;:\;\exists\lar{f_\lambda}\;:\:
{\cal K}\lar{f_\lambda}=\lar{f_\lambda}\lambda
\right\}
\,.
\end{eqnarray}
The continuous spectrum contains all numbers $\lambda$ where
$(\lambda-{\cal K})^{-1}$ formally exists, but is not bounded 
\begin{eqnarray}
\sigma_{\rm c}=\left\{
\lambda\in{\mathbb C}\backslash\sigma_{\rm p}\;:\;
||(\lambda-{\cal K})^{-1}||=\infty
\right\}
\,.
\end{eqnarray}
The discrete spectrum $\sigma_{\rm d}$ is that part of the point spectrum 
$\sigma_{\rm p}$ which incorporates all isolated points $\lambda$
of $\sigma_{\rm p}$ with a finite number of corresponding 
eigenfunctions $\lar{f_\lambda}$.
The continuous spectrum $\sigma_{\rm c}$ may also be divided into two 
parts, the absolute continuous spectrum $\sigma_{\rm ac}$, where
$d{\cal E}(\lambda)/d\lambda$ exists as a weakly integrable operator,
and the remaining singular continuous spectrum $\sigma_{\rm sc}$.

To provide some physical insight into these abstract quantities we shall 
investigate the spectrum for a few examples.
The discrete spectrum $\sigma_{\rm d}$ describes localised states, 
such as bound states or states of a field confined in a finite volume.
The point spectrum $\sigma_{\rm p}$ may contain more points with 
additional characteristics. 
E.g., if the operator $\cal K$ governs the dynamics of the Maxwell field 
$A_\mu$ there is an infinite set of eigenfunctions at the point 
$\lambda=0$. 
These functions correspond to the gauge invariance of this theory and do not 
change physical quantities.
The absolute continuous spectrum $\sigma_{\rm ac}$ represents usually the 
scattering states, but the singular continuous spectrum $\sigma_{\rm sc}$
may be related to more strange phenomena, like quasi-bound states, scattering 
states in average, fractal measure $d\mu_\wp$ 
(cf. Sec. \ref{stheorem} below), chaotic behaviour, etc.

Fortunately, for smooth and regular coefficients $g_{\mu\nu}$ with 
an appropriate asymptotic behaviour the spectrum of the $\cal K$-operator 
is either purely discrete $\sigma({\cal K})=\sigma_{\rm d}$
(for a finite volume) or absolute continuous 
$\sigma({\cal K})=\sigma_{\rm ac}$  
(for an infinite volume, see e.g. \cite{reed+simon}).

\subsection{Spectral theorem}\label{stheorem}

For our main intention, the diagonalisation of the Lagrange function, 
it is suitable to make use of the following theorem:
For every self-adjoint operator ${\cal K}$ acting on a separable Hilbert 
space there exists a unitary transformation ${\cal U}$ which diagonalises 
it : ${\cal U}{\cal K}\,{\cal U}^\dagger ={\cal M}$.
${\cal M}$ denotes the multiplication by argument: 
$({\cal M}f)(\lambda)=\lambda f(\lambda)$.
Because $\cal K$ is $\mathbb C$-real, i.e. $({\cal K}\phi)^*={\cal K}(\phi^*)$, 
we may construct a quasi-unitary transformation
\begin{eqnarray}
{\cal V}=\left(
\begin{array}{l}
\Re({\cal U})\\ 
\Im({\cal U})
\end{array}
\right)\; : \; L_2(G,g^{00})
\;\longrightarrow\;
\bigoplus\limits_\wp\, 
L_2(\sigma({\cal K}),\mu_\wp)\stackrel{\rm def}{=}
L_2(\sigma,{\cal V})
\,,
\end{eqnarray}
which is $\mathbb C$-real $({\cal V}\phi)^*={\cal V}(\phi^*)$ and does also 
diagonalise the operator ${\cal V}{\cal K}{\cal V}^\dagger ={\cal M}$.  

Accordingly, the Hilbert space $L_2(\sigma,{\cal V})$ is restricted to 
real numbers and the associated scalar product reads  
\begin{eqnarray}
\label{scalarV}
\scalar{u}{v}_\sigma=
\scalar{{\cal V}^\dagger u}{{\cal V}^\dagger v}_1=
\sum\limits_\wp 
\int\limits_{\sigma({\cal K})} d\mu_\wp(\lambda)\,
u_\wp(\lambda)\,v_\wp(\lambda)
\stackrel{\rm def}{=}\sumint{\Gamma}u_\Gamma v_\Gamma
\,.
\end{eqnarray}
Because $L_2(\sigma,{\cal V})$ is a real Hilbert space over $\mathbb R$, 
the usual complex conjugation of the first argument in the scalar product 
disappears. 

For a discrete spectrum $\sigma({\cal K})=\sigma_{\rm d}$ the measure 
$d\mu_\wp(\lambda)$ denotes simply a sum and for an absolute continuous 
spectrum $\sigma({\cal K})=\sigma_{\rm ac}$ an elementary integral 
possibly together with a $\wp$-summation, cf. Eq. (\ref{scalarV}).  
For example, the $\wp$-sum may describe the angular quantum numbers for the 
Laplacian in spherical coordinates $\wp=\ell,m$.
Both, the $\wp$-summation and the integration with the measure 
$d\mu_\wp(\lambda)$ are now abbreviated by the index $\Gamma$.  
    
Performing the transformation of the fields 
$\lar{Q}_\sigma={\cal V}\lar{\Phi}_1$ 
the Lagrange function can be diagonalised
\begin{eqnarray}
\label{diagL}
L=\frac{1}{2}\scalar{\dot Q}{\dot Q}_\sigma-
\frac{1}{2}\sca{Q}{\cal M}\lar{Q}_\sigma=
\frac{1}{2}\sumint{\Gamma}
\left(
\dot Q^2_\Gamma-\omega^2_\Gamma Q^2_\Gamma
\right)
\,,
\end{eqnarray}
with $\omega^2_\Gamma\stackrel{\rm def}{=}\lambda_\Gamma
\in\sigma({\cal K})\subseteq{\mathbb R}_+$
which will be called eigenfrequencies.

One should note that the $\lar{Q}_\sigma$ still depend on time
$\lar{Q(t)}_\sigma$, only the spatial dependence is transformed 
by $\cal V$. 
Owing to the reality of the transformation $\cal V$ the amplitudes 
$Q_\Gamma(t)$ are real as well.

\subsection{Canonical quantisation}\label{canon}  

Starting with the diagonal Lagrange function in Eq. (\ref{diagL}) we are  
able to perform the canonical quantisation procedure by imposing the usual 
equal time commutation relations
\begin{eqnarray}
\left[\scalar{u}{\hat Q(t)}_\sigma,
\scalar{\hat P(t)}{v}_\sigma\right]
&=&
i\scalar{u}{v}_\sigma
\,,
\nonumber\\
\left[\scalar{u}{\hat Q(t)}_\sigma,
\scalar{\hat Q(t)}{v}_\sigma\right]
&=&
\left[\scalar{u}{\hat P(t)}_\sigma,
\scalar{\hat P(t)}{v}_\sigma\right]=0
\,,
\end{eqnarray}
which hold for all $\lar{u}_\sigma$ and $\lar{v}_\sigma$.
In this representation the canonical conjugated momenta are simply 
determined by $\lar{P}_\sigma=\lar{dQ/dt}_\sigma$.

Due to the isometry of the transformation $\cal V$ these commutation 
relations are completely equivalent to the corresponding relations for the 
field $\hat\Phi$.
For a static metric the inner product is related to the scalar product via
\begin{eqnarray}
\inner{\psi}{\phi}=i\scalar{\psi}{\dot\phi}_1
-i\scalar{\dot\psi}{\phi}_1
\,.
\end{eqnarray}
As a consequence, the relations above are indeed identical to the commutators 
of the fields
\begin{eqnarray}
\left[\inner{\psi}{\hat\Phi},
\inner{\hat\Phi}{\phi}\right]=
\inner{\psi}{\phi}
\,.
\end{eqnarray}
The Hamiltonian splits up into an infinite set of commuting parts describing
harmonic oscillators that are appropriate for a particle definition
\begin{eqnarray}
\label{HamPQ}
\hat H=\frac{1}{2}\scalar{\hat P}{\hat P}_\sigma+
\frac{1}{2}\sca{\hat Q}{\cal M}\lar{\hat Q}_\sigma=
\frac{1}{2}\sumint{\Gamma}
\left(
\hat P^2_\Gamma+\omega^2_\Gamma\hat Q^2_\Gamma
\right)
\,.
\end{eqnarray}
In terms of the creators $\lar{\hat A^\dagger }$ and annihilators 
\begin{eqnarray}
\lar{\hat A}=\frac{1}{\sqrt{2}}\left(
{\cal M}^{1/4}\lar{\hat Q(t=0)}+i{\cal M}^{-1/4}\lar{\hat P(t=0)}\right)
\end{eqnarray}
the Hamiltonian can be cast into the form
\begin{eqnarray}
\label{HamA}
\hat H=\frac{1}{2}\sca{\hat A^\dagger }{\cal M}^{1/2}\lar{\hat A}_\sigma+
\frac{1}{2}\sca{\hat A}{\cal M}^{1/2}\lar{\hat A^\dagger }_\sigma=
\sumint{\Gamma}\frac{\omega_\Gamma}{2}
\left(\hat A_\Gamma^\dagger \hat A_\Gamma+
\hat A_\Gamma\hat A_\Gamma^\dagger \right)
\,.
\end{eqnarray}
For a discrete spectrum $\sigma({\cal K})=\sigma_{\rm d}$ this already 
defines the physical particles because we have now creation and annihilation 
operators $\hat A_\Gamma^\dagger$ and $\hat A_\Gamma$ that diagonalise 
the Hamiltonian (which is also the energy operator).
  
For a continuous spectrum $\sigma({\cal K})=\sigma_{\rm c}$ the 
quantities $\hat P_\Gamma$, $\hat Q_\Gamma$, $\hat A_\Gamma$ and 
$\hat A_\Gamma^\dagger $ are not well-defined operators but operator-valued 
distributions because
\begin{eqnarray}
\left[\hat Q_\Gamma(t),\hat P_\Lambda(t)\right]=
i\delta(\Gamma,\Lambda)
\,,
\end{eqnarray}
which is a Dirac $\delta$-distribution for continuous indices 
$\Gamma,\Lambda$. 
But a product of two distributions acting on the same linear space
(e.g. the Schwartz/Sobolev space ${\mathfrak S}_1$ 
for the $\delta$-distribution),
i.e. with the same index $\Gamma$, is not well-defined. 
This reflects the infinite-volume divergence in quantum field theory. 
Consequently, the Hamiltonian in Eqs. (\ref{HamPQ}) and (\ref{HamA}) 
is not well-defined. 
It may only be viewed as a formal expression until an appropriate
regularisation method has been applied.

\subsection{Vacuum definition}\label{vacuum}

In order to get rid of the singularities discussed above and to obtain
well-defined operators $\hat a_I$ we introduce a complete orthonormal
and real basis $\lar{b_I}_\sigma$ with $I\in\mathbb N$ of the
separable Hilbert space $L_2(\sigma,{\cal V})$ and define
\begin{eqnarray}
\hat a_I\stackrel{\rm def}{=}\scalar{\hat A}{b_I}_\sigma
\,.
\end{eqnarray}
For a discrete spectrum $\sigma({\cal K})=\sigma_{\rm d}$ we may choose 
$b_I(\Gamma)=\delta_{\Gamma I}$ which leads us back to the operators 
$\hat A_\Gamma$.
For a continuous spectrum $\sigma({\cal K})=\sigma_{\rm c}$ this 
coincidence does not hold.
Due to $\scalar{b_I}{b_J}_\sigma=\delta_{IJ}$ with a 
Kronecker-$\delta_{IJ}$ the $\hat a_I$ are well-defined operators 
with $[\hat a_I,\hat a_J^\dagger ]=\delta_{IJ}$ instead of 
operator-valued distributions with 
$[\hat A_\Gamma,\hat A_\Lambda^\dagger ]=\delta(\Gamma,\Lambda)$.
Unfortunately, for a continuous spectrum $\sigma({\cal K})=\sigma_{\rm c}$
the operators $\hat a_I$ are now well-defined, but do not {\em exactly} 
diagonalise the Hamiltonian. 
But -- as we shall see later in Section \ref{eigen} -- one may choose an 
appropriate basis $\lar{b_I}_\sigma$ for which the operators 
$\hat a_I$ {\em approximately} diagonalise the Hamiltonian.

The corresponding number operators take the usual form 
$\hat n_I\stackrel{\rm def}{=}\hat a_I^\dagger \hat a_I$.
The Fock space $\mathfrak F$ which contains all pure states $\ket{\Psi}$
of the quantum field $\hat\Phi$ is now {\em defined} as the completion of the 
linear hull of the proper eigenvectors of these commuting operators 
$\hat n_I$ for all indices $I\in\mathbb N$
\begin{eqnarray}
\label{Fock}
{\mathfrak F}\stackrel{\rm def}{=}\overline{\stackrel{ }{\rm lin}\left\{
\ket{\Psi}\,:\,\forall_I\,:\,\hat n_I\ket{\Psi}=\ket{\Psi}n_I
\right\}}
\,.
\end{eqnarray}
As a consequence, the spectrum of the operators $\hat n_I$ in this Fock 
space is a pure point spectrum $\sigma(\hat n_I)=\sigma_{\rm p}$. 
With the same arguments as already used for the quantisation of the harmonic 
oscillator the commutation relations 
$[\hat a_I,\hat a_J^\dagger ]=\delta_{IJ}$ imply
$\sigma(\hat n_I)=\mathbb N$.   
It should be mentioned here that this definition of the Fock space 
$\mathfrak F$ is slightly different to the frequently employed approach based 
on the one-particle Hilbert space 
${\mathfrak H}\simeq L_2^{\mathbb C}(G,g^{00})$ 
(see e.g. \cite{wald-p,reed+simon,ashtekar,kay,dimock}) 
\begin{eqnarray}
{\mathfrak F}={\mathbb C}
\oplus{\mathfrak H}\oplus
\left({\mathfrak H}\otimes{\mathfrak H}\right)_{\rm symm}\oplus
\left({\mathfrak H}\otimes{\mathfrak H}
\otimes{\mathfrak H}\right)_{\rm symm}
\oplus\dots
\,.
\end{eqnarray}
Nevertheless, these distinct definitions are related if one divides the Fock 
space in Eq. (\ref{Fock}) into orthogonal subspaces labelled by different
values of the total number of particles 
$n_{\rm total}\stackrel{\rm def}{=}\sum_I n_I\in\mathbb N$.
 
Accordingly, the vacuum is defined as the eigenvector of all commuting 
operators $\hat n_I$ with eigenvalue zero
\begin{eqnarray}
\forall_I\;:\;\hat n_I\ket{0}=0\quad{\rm i.e.}\quad\hat a_I\ket{0}=0
\,.
\end{eqnarray}
This definition is independent of the special choice of the basis 
$\lar{b_I}_\sigma$.
To prove this statement, we use the completeness of the basis 
$\lar{b_I}_\sigma$ to obtain the following result
\begin{eqnarray}
\forall\;\lar{\zeta}_\sigma\;:\;
\scalar{\hat A}{\zeta}_\sigma\ket{0}=0
\,.
\end{eqnarray}
If we regularise the formal expression for the Hamiltonian in Eq. (\ref{HamA})
via insertion of a complete basis (in principle together with a convergence 
factor, $\exp(-I\varepsilon)$ for instance)
\begin{eqnarray}
\label{regHam}
\hat H
&=&
\frac{1}{2}\sum\limits_I
\left(
\scalar{{\cal M}^{1/4}\hat A^\dagger }{b_I}_\sigma
\scalar{b_I}{{\cal M}^{1/4}\hat A}_\sigma+
\scalar{{\cal M}^{1/4}\hat A}{b_I}_\sigma
\scalar{b_I}{{\cal M}^{1/4}\hat A^\dagger }_\sigma
\right)
\nonumber\\
&=&
\sum\limits_I
\left(
\scalar{\hat A^\dagger }{{\cal M}^{1/4}b_I}_\sigma
\scalar{{\cal M}^{1/4}b_I}{\hat A}_\sigma+
\frac{1}{2}
\scalar{{\cal M}^{1/4}b_I}{{\cal M}^{1/4}b_I}_\sigma
\right)
\,,
\end{eqnarray}
it appears as a divergent sum of some non-negative operators of the structure 
$\hat X^\dagger _I\hat X_I$ and remaining $\mathbb C$-numbers.
The above defined vacuum is the ground state of all operators
$\hat X^\dagger _I\hat X_I$ and in this regard also the ground state
of the Hamiltonian.
Hence, the divergent amount of $\mathbb C$-number terms represents the 
zero-point energy.
The infinite summation over the index $I$ corresponds to the sum over 
arbitrary high frequencies and -- for a continuous spectrum -- the summation 
of an infinite number of basis elements for a given frequency interval.
The first infinity, the infinite energy divergence, is always present in 
quantum field theory and the latter, the infinite volume divergence, only 
for non-discrete spectra.
   
In the Minkowski space-time the above defined vacuum coincides (of course) with 
the usual Minkowski vacuum $\ket{0}=\ket{0_{\rm M}}$. 
In the Schwarzschild space-time this state -- which is the ground state of the 
Hamiltonian -- is called the Boulware \cite{boulware} state 
$\ket{0}=\ket{\Psi_{\rm B}}$. 

The particle definition presented above can be reproduced utilising the 
well-known approach based on the inner product:
The basis elements $\lar{b_I}_\sigma$ of the Hilbert space
$L_2(\sigma,{\cal V})$ are normalised and therefore correspond to 
functions  $e_i(\f{r})$ via $\lar{e_I}_1=
{\cal V}^\dagger \lar{b_I}_\sigma$ which are also normalised 
$\scalar{e_I}{e_J}_1=\delta_{IJ}$ and build up a basis of 
the Hilbert space $L_2(G,g^{00})$.
As a consequence, the operators $\hat a_I$ correspond to localised 
wave-packets $\lar{F_I(t)}_1$ which are defined as follows
\begin{eqnarray}
\lar{F_I(t)}_1=\left(4{\cal K}\right)^{-1/4}
\exp\left(-i{\cal K}^{1/2}t\right)
\lar{e_I}_1
\,.
\end{eqnarray}
These quantities are solutions of the Klein-Fock-Gordon equation
$\partial^2_t\lar{F_I}_1=-{\cal K}\lar{F_I}_1$
and normalised with respect to the inner product
\begin{eqnarray}
\inner{F_I}{F_J}=-\inner{F_I^*}{F_J^*}=\delta_{IJ}
\quad,\quad
\inner{F_I^*}{F_J}=\inner{F_I}{F_J^*}=0
\,.
\end{eqnarray}
Comparison with the particle definition via the inner product verifies indeed 
the identification 
\begin{eqnarray}
\hat a_I=\inner{F_I}{\hat\Phi}
\,.
\end{eqnarray}
The functions $F_I$ and $F_I^*$ form a complete set of solutions of the 
Klein-Fock-Gordon equation. 
Hence, the field $\hat\Phi$ may be expanded via
\begin{eqnarray}
\hat\Phi=\sum\limits_I\hat a_IF_I+\hat a_I^\dagger F_I^*
\,,
\end{eqnarray}
which demonstrates again the equivalence of the approaches.

\subsection{Eigenfunctions}\label{eigen}

For a point spectrum $\sigma_{\rm p}$ there exist proper eigenfunctions 
$f_\Gamma\in L_2(G,g^{00})$ with
${\cal K}\lar{f_\Gamma}_1=\omega_\Gamma^2\lar{f_\Gamma}_1$, 
but for a continuous spectrum $\sigma_{\rm c}$ this is of course not the case.
Nevertheless, it is in many cases possible to find an analogue. 
If pointwise defined functions $f_\Gamma(\f{r})$ 
(or -- more generally -- locally integrable functions 
$f_\Gamma\in L_1^{\rm local}$)
exist such that
\begin{eqnarray}
\scalar{\zeta}{{\cal V}\phi}_\sigma=
\sumint{\Gamma}\zeta_\Gamma 
\int\limits_G d^3\f{r} 
f_\Gamma(\f{r})\,\phi(\f{r})
\end{eqnarray}
holds for all $\phi\in C^\infty_0(G)$ and $\zeta\in C^\infty_0(\sigma)$ the 
functions $f_\Gamma(\f{r})$ are called (generalised) eigenfunctions of 
the $\cal K$-operator.
In contrast to the proper ($\sigma_{\rm p}$) eigenfunctions with
$f_\Gamma\in{\rm Def}({\cal K})\subset L_2(G,g^{00})$ the generalised 
($\sigma_{\rm c}$) eigenfunctions do not belong to the Hilbert space 
$f_\Gamma\not\in L_2(G,g^{00})$ and (of course) also not to
the domain of definition of the ${\cal K}$-operator 
$f_\Gamma\not\in{\rm Def}({\cal K})$.
However, due to $\cal VK=MV$ also the generalised eigenfunctions fulfil the 
pointwise/local (generalised) eigenvalue equation 
${\cal K}_{\rm local}f_\Gamma(\f{r})=
\omega_\Gamma^2f_\Gamma(\f{r})$.
This is a very important relation for the calculation of these eigenfunctions.
If the (generalised) eigenfunctions exist, the transformation of the fields 
$\lar{\Phi}_1={\cal V}^\dagger \lar{Q}_\sigma$ 
can be described by the pointwise/local identity
\begin{eqnarray}
\hat\Phi(\f{r},t)
=\sumint{\Gamma}\hat Q_\Gamma(t)\,f_\Gamma(\f{r})
\,.
\end{eqnarray}
Even though the generalised eigenfunctions are not in $L_2(G,g^{00})$, 
they may be thought as a (singular) limiting case of 
$L_2(G,g^{00})$-functions:
In the following considerations we assume 
$d\mu(\lambda)=d\lambda$ and $\sigma=\mathbb R$ for reasons of simplicity.
The $L_2(\sigma,{\cal V})$-basis functions $b_I(\lambda)$ can be 
squeezed and translated
$b_{I(\Gamma)}^\varepsilon(\lambda)=b_I
(\lambda/\varepsilon-\lambda_\Gamma/\varepsilon)/\sqrt{\varepsilon}$
and are still a basis of $L_2(\sigma,{\cal V})$.
Evaluating the (singular) limiting case of these squeezed basis functions
\begin{eqnarray}
\lim\limits_{\varepsilon\downarrow0}
\frac{b_{I(\Gamma)}^\varepsilon(\lambda)}{\sqrt{\varepsilon}}
=\lim\limits_{\varepsilon\downarrow0}
\frac{b_I(\lambda\,\varepsilon^{-1}-\lambda_\Gamma\,\varepsilon^{-1})}
{\varepsilon}
={\cal N}_{I(\Gamma)}\;\delta(\lambda-\lambda_\Gamma)\,,
\end{eqnarray}
where ${\cal N}_{I(\Gamma)}$ denotes some normalisation factor,
we observe that every generalised eigenfunction $f_\Gamma(\f{r})$ can be 
locally approximated by appropriately chosen wave packets
$\lar{e_{I(\Gamma)}^\varepsilon}_1=
{\cal V}^\dagger \lar{b_{I(\Gamma)}^\varepsilon}_\sigma$
\begin{eqnarray}
\lim\limits_{\varepsilon\downarrow0}
\frac{e_{I(\Gamma)}^\varepsilon(\f{r})}{\sqrt{\varepsilon}}
={\cal N}_{I(\Gamma)}\;f_\Gamma(\f{r})
\,.
\end{eqnarray}
Accordingly, also the operator-valued distributions $\hat A_\Gamma$ may 
be considered as a singular limiting case of the regular operators 
$\hat a^{\varepsilon}_{I(\Gamma)}$ 
\begin{eqnarray}
\lim\limits_{\varepsilon\downarrow0}
\frac{\hat a_{I(\Gamma)}^\varepsilon}{\sqrt{\varepsilon}}
={\cal N}_{I(\Gamma)}\;\hat A_\Gamma
\,.
\end{eqnarray}
The divergent factor $1/\sqrt{\varepsilon}$ indicates
the singular character of the generalised eigenfunctions 
(e.g. plane waves) in contrast to the regular basis elements
(wave packets).
Of course, in realistic experiments one never deals with plane waves, 
but wave packets.
On the other hand, the calculations with plane waves are usually much simpler.
Hence, in the following we shall perform our evaluations with eigenfunctions 
always bearing in mind their character as a singular limiting case of 
regular objects.  

\subsection{Continuum normalisation}\label{continuum}

To investigate the physical consequences caused by the singular behaviour of 
the product of two distributions
$\hat N_\Gamma=\hat A^\dagger _\Gamma\hat A_\Gamma$
-- expressed by the factor $1/{\varepsilon}$ --
we consider a quantum field confined in a finite volume $V$ and study the 
limiting case $V\rightarrow\infty$.
This limit may be interpreted as the transition from a discrete spectrum 
$\sigma({\cal K})=\sigma_{\rm d}$ to a continuous one 
$\sigma({\cal K})=\sigma_{\rm c}$.
For a 3-dimensional cubic volume $V$ the indices $\Gamma$ correspond, 
for example, to discrete wave-numbers $\f{k}$.
In the continuum limit $V\rightarrow\infty$ the $\f{k}$-sum transforms into 
an integral over $d^3\f{k}$ via 
\begin{eqnarray}
\sum\limits_{\f{k}}\rightarrow{\cal N}_V V\int d^3\f{k}
\,.
\end{eqnarray}
${\cal N}_V$ denotes a normalisation factor which depends on the imposed 
boundary conditions (Dirichlet, Neumann, periodic, etc.)
and the shape of the domain $G$.
The Kronecker-$\delta_{\f{k},\f{k}'}$ converts into a 
Dirac-$\delta^3(\f{k}-\f{k}')$ in an analogue way
${\cal N}_V V\delta_{\f{k},\f{k}'}\rightarrow\delta^3(\f{k}-\f{k}')$.
Ergo, the singularity $\delta^3(\f{k}-\f{k})$ displays the infinite-volume 
divergence $\delta(\f{k},\f{k})=\delta^3(\f{k}-\f{k})={\cal N}_V V$.
Recalling the formal expression for the Hamiltonian in Eq. (\ref{HamA})
\begin{eqnarray}
\hat H
=
\sumint{\Gamma}\omega_\Gamma
\left(\hat A_\Gamma^\dagger \hat A_\Gamma+
\frac{1}{2}\delta(\Gamma,\Gamma)\right)
=
\sumint{\Gamma}
\left(
\omega_\Gamma\hat N_\Gamma+
\frac{\omega_\Gamma}{2}\delta(\Gamma,\Gamma)
\right)
\,,
\end{eqnarray}
we observe that 
-- in addition to the mode summation/integration --
its indefinite character exactly exhibits this divergence.
Indeed, if one examines the continuum limit of the Hamiltonian
\begin{eqnarray}
\hat H^{\rm d}=\sum\limits_{\f{k}}
\left(\hat N_{\f{k}}^{\rm d}+\frac{1}{2}\right)|\f{k}|
\longrightarrow
\hat H^{\rm c}=\int d^3\f{k}
\left(\hat N_{\f{k}}^{\rm c}+\frac{1}{2}\delta^3(\f{k}-\f{k})\right)|\f{k}|
\,,
\end{eqnarray}
the limiting number "operator" $\hat N_{\f{k}}^{\rm c}$ can be identified via 
${\cal N}_V V\hat N_{\f{k}}^{\rm d}\rightarrow\hat N_{\f{k}}^{\rm c}$.
The singular character of this formal expression may be exemplified with the 
following consideration: 
If the state of the quantum field corresponds to thermal equilibrium at a 
temperature $T>0$, the expectation value of the number operator 
$\hat N_{\f{k}}^{\rm d}$ equals the Bose-Einstein distribution for arbitrary 
large but finite volumes $V$. 
For an infinite volume the expectation value of the quantity 
$\hat N_{\f{k}}^{\rm c}$ diverges owing to the factor ${\cal N}_V V$.
The expectation values of the regular operators $\hat n_I$ are (of course) 
still finite and behave as the Bose-Einstein distribution evaluated at some 
averaged frequency $\omega_I$.
 
\subsection{Bogoliubov coefficients}\label{bogol}

So far we have considered static space-times and developed an appropriate 
particle definition.
If we now drop the restriction to stationary metrics and take dynamical 
space-times into account, the question concerning particle creation arises.
A variation of the metric $g_{\mu\nu}$ induces a change of the 
$\cal K$-operator and -- possibly -- the corresponding Hilbert space
$L_2(G,g^{00})$.
A function, which belongs initially to $L_2(G^{\rm in},g^{00}_{\rm in})$
may be later (e.g. if a horizon has formed) not in 
$L_2(G^{\rm out},g^{00}_{\rm out})$ but a distribution with respect to 
the $L_2(G^{\rm out},g^{00}_{\rm out})$-scalar product.
As a consequence, it is not clear whether the Bogoliubov coefficient 
(see e.g. \cite{birrell}--\cite{wipf}) describing the particle creation
\begin{eqnarray}
\beta_{IJ}=\inner{F_I^{*}}{F_J}
\end{eqnarray}
exists for all $F_I\in L_2(G^{\rm in},g^{00}_{\rm in})$ and 
$F_J\in L_2(G^{\rm out},g^{00}_{\rm out})$ or not.
However, $C^\infty_0$-functions belong to the domain of definition of all 
(tempered) distributions.
Thus for $F^{\rm in}_I\in C^\infty_0(G^{\rm in})$ and
$F^{\rm out}_J\in C^\infty_0(G^{\rm out})$ the Bogoliubov coefficients always 
exist provided the metric can be cast into an analytic form.
Similar to the previous Sections all other quantities
(e.g. generalised eigensolutions) have to be approximated with 
$C^\infty_0$-functions.

\section{Flat space-time examples}\label{flat}

In the previous Section we have derived a canonical definition of particles 
for curved space-times which fulfil certain conditions.
In the following we are going to apply this approach to the most simple 
example of a flat space-time in order to achieve a deeper insight into the 
physical consequences of the used mathematical theorems.

For the unbounded 1+1 dimensional Minkowski space-time with $ds^2=dt^2-dx^2$ 
the $\cal K$-operator reads
\begin{eqnarray}
\label{einK}
{\cal K}=-\frac{\partial^2}{\partial x^2}
\,,
\end{eqnarray}
together with the domain $G=(-\infty<x<\infty)$. 
The infinite volume of this domain and the regularity of the metric cause a 
purely absolute continuous spectrum 
$\sigma({\cal K})=\sigma_{\rm ac}={\mathbb R}_+$.
The unitary transformation ${\cal U}$ (see Section \ref{stheorem}) is simply 
the one-dimensional Fourier transformation ${\cal U}={\cal F}$.
The quasi-unitary transformation ${\cal V}$ takes the real and imaginary 
parts separately leading to the generalised eigenfunctions $\sin\omega x$ 
and $\cos\omega x$. 
Hence, the expansion of the field $\hat\Phi$ takes the following form
\begin{eqnarray}
\hat\Phi(x,t)
&=&
\sumint{\omega}
\hat Q_{\omega,\rm c}(t)\cos(\omega x)+
\hat Q_{\omega,\rm s}(t)\sin(\omega x)
\nonumber\\
&=&
{\cal N}_\mu\int\limits_0^\infty \frac{d\omega}{\sqrt{2\omega}}\,
\left(
\hat A^\dagger _{\omega,\rm c}\,e^{i\omega t}\,\cos(\omega x)+
\hat A^\dagger _{\omega,\rm s}\,e^{i\omega t}\,\sin(\omega x)+
{\rm h.c.}
\right)
\,,
\end{eqnarray}
with a normalisation factor ${\cal N}_\mu$ depending on the explicit
form of the measure $d\mu(\omega^2)$, e.g. $d\mu(\omega^2)=d\omega$ or 
$d\mu(\omega^2)=d\omega/2\pi$ etc.
The spectrum of the $\cal K$-operator discussed above is twice degenerated, 
i.e. there are two independent generalised eigenfunctions 
($\sin\omega x$ and $\cos\omega x$) for every point $\lambda=\omega^2$ 
of the spectrum $\sigma$.
This degeneracy of the spectrum allows for the definition of particles with 
a definite direction of propagation:
With a simple linear transformation we may rearrange the expansion of the 
field  
\begin{eqnarray}
\hat\Phi(x,t)
=
{\cal N}_\mu\int\limits_0^\infty \frac{d\omega}{\sqrt{4\omega}}\,
\left(
\hat A^\dagger _{\omega,+}\,e^{i\omega t+i\omega x}+
\hat A^\dagger _{\omega,-}\,e^{i\omega t-i\omega x}+
{\rm h.c.}
\right)
\,.
\end{eqnarray}
The new introduced quantities $\hat A_{\omega,+}$, 
$\hat A_{\omega,-}$, $\hat A^\dagger _{\omega,+}$ and 
$\hat A^\dagger _{\omega,-}$ obey the same commutation relations as the 
original ones $\hat A_{\omega,\rm c}$, $\hat A_{\omega,\rm s}$, 
$\hat A^\dagger _{\omega,\rm c}$ and $\hat A^\dagger _{\omega,\rm s}$. 
Thus they also describe particles. 
In contrast to the original particles which correspond to standing waves with 
different phases ($\sin\omega x$ and $\cos\omega x$) the new particles 
describe left-moving and right-moving waves according to $\exp(\pm i\omega x)$.
These complex functions are suitable for a definition of particles
$\hat A_\Gamma$ but do not correspond to Hermitian amplitudes 
$\hat Q_\Gamma$.

As a second example we study the situation of a bounded domain 
$G=(0<x<\infty)$ in a 1+1 dimensional Minkowski space-time with a Dirichlet 
boundary condition (a mirror) at $x=0$. 
Even though $\cal K$ seems to have the same form as in Eq. (\ref{einK}) 
it denotes a different operator as a result of the boundary condition.
With the same arguments the spectrum is purely absolute continuous
$\sigma=\sigma_{\rm ac}={\mathbb R}_+$.
In contrast to the previous example this spectrum is not degenerated. 
Every point $\lambda=\omega^2$ of $\sigma$ corresponds to exactly {\em one} 
generalised eigenfunction, i.e. $\sin\omega x$.
As a consequence, the definition of particles with a certain direction
(left-moving or right-moving) is {\em not} possible.
This result is physical reasonable if one takes conservation laws into 
account. 
Every left-moving component will be reflected by the mirror at $x=0$ after 
some period of time and turns its direction into right-moving and vice
versa.

A finite domain $G=(0<x<L)$ in a 1+1 dimensional Minkowski space-time with 
Dirichlet boundary conditions at $x=0$ and $x=L$ of course possesses a purely 
discrete spectrum $\sigma=\sigma_{\rm d}$ with proper eigenfunctions
$\sim\sin({\mathbb N}\,\pi x/L)$.
The insertion of mirrors represented by Dirichlet boundary conditions usually 
lowers the "density" of the spectrum $\sigma$, i.e. the number of 
eigenfunctions. 

\subsection{Ingoing and outgoing particles}\label{ingoing}

Now we shall extend our investigations to the 3+1 dimensional Minkowski 
space-time described by different coordinate systems.
Using spherical coordinates $r,\vartheta,\varphi$ it will turn out that the 
definition of ingoing or outgoing particles is {\em not} possible within the 
canonical approach. 
This is a consequence of the spectral properties of the operator
\begin{eqnarray}
{\cal K}=-\nabla^2=-\frac{\partial^2}{\partial\f{r}^2}
\,,
\end{eqnarray}
together with the domain $G={\mathbb R}^3$.
Expressed by Cartesian coordinates $\f{r}=(x,y,z)^T$ the generalised 
eigenfunctions take the simple form $\sin(\f{k}\f{r})$ and $\cos(\f{k}\f{r})$
with $|\f{k}|=\omega$.
As it is well-known these functions form a complete basis of 
$L_2({\mathbb R}^3)$.

Employing spherical coordinates $r,\vartheta,\varphi$ the Cartesian 
eigenfunctions can be expanded with the aid of the equality
\begin{eqnarray}
\exp(i\f{k}\f{r})=\sum\limits_{\ell=0}^{\infty}
i^\ell(2\ell+1)j_\ell(\omega r)P_\ell(\cos\theta)
\,,
\end{eqnarray}
with $\f{k}\f{r}=\omega r\cos\theta$ and the Legendre polynomials 
$P_\ell$.
By inspection we recognise the following fact:
For a given angular behaviour the  spherical Bessel functions 
$j_\ell(\omega r)$ are already complete to describe the radial dependence.
The Neumann $n_\ell(\omega r)$ or Hankel functions $h_\ell^\pm(\omega r)$
are not required and would be "over-complete".
Therefore they do not describe additional degrees of freedom and do not enter 
the particle definition. 
This result can also be derived by considering the spectrum of the 
$\cal K$-operator.
Due to the singular behaviour of the functions 
$n_\ell(\omega r)$ and $h_\ell^\pm(\omega r)$ at $r=0$
they are not eigenfunctions.

To acquire real eigenfunctions we have to introduce redefined spherical 
harmonics 
${\cal Y}_{\ell m}(\vartheta,\varphi)\stackrel{\rm def}{=}{\cal N}_{\ell}
P^m_\ell(\cos\vartheta)\cos(m\varphi)$ for $m\geq0$ and
${\cal Y}_{\ell m}(\vartheta,\varphi)\stackrel{\rm def}{=}{\cal N}_{\ell}
P^m_\ell(\cos\vartheta)\sin(m\varphi)$ for $m<0$.
$P^m_\ell$ denote the associated Legendre polynomials and 
${\cal N}_{\ell}$ are normalisation factors.
Accordingly, the complete set of real and orthogonal eigenfunctions reads
\begin{eqnarray}
f_{\omega\ell m}(r,\vartheta,\varphi)=
{\cal N}_{\omega\ell}\,j_\ell(\omega r)
{\cal Y}_{\ell m}(\vartheta,\varphi)
\,.
\end{eqnarray}
Again we observe the occurrence of exactly {\em one} eigenfunction 
$f_{\omega\ell m}$ per eigenfrequency $\omega$ for a fixed angular 
dependence $\ell,m$. 
The regularity at $r=0$ plays the role of an effective boundary condition
and forbids the existence of additional eigenfunctions such as
$n_\ell(\omega r)$ or $h_\ell^\pm(\omega r)$.
As a consequence, within the canonical approach it is {\em not} possible to 
define radial ingoing or outgoing particles in the Minkowski space-time. 
Functions like $\exp(\pm i\omega r)/r$ are {\em not} eigenfunctions of the 
Laplacian and therefore {\em not} solutions of the wave equation
\begin{eqnarray}
\Box\frac{\exp(i\omega t\pm i\omega r)}{r}=
-4\pi e^{i\omega t}\delta^3(\f{r})\neq 0
\,.
\end{eqnarray}
Expanding the field $\hat\Phi$ into functions that do not satisfy the 
equation of motion $\Box\hat\Phi=0$ would abandon the independence of the 
distinct particles.
Functions like $\exp(\pm i\omega r)/r$ correspond to the resolvents 
${\cal R}(\omega^2\pm i\varepsilon)$ of the operator ${\cal K}$ 
(remember $\sigma({\cal K})\subset\mathbb R$).
Particles are defined with respect to the eigenfunctions which are 
representations of the spectral family ${\cal E}$ of the operator ${\cal K}$. 
Into this spectral family ${\cal E}$ the resolvents themselves do not enter, 
but linear combinations of them:  
${\cal R}(\lambda-i\varepsilon)-{\cal R}(\lambda+i\varepsilon)
\rightarrow{\cal E}(\lambda)$ which again leads to 
$\sin(\omega r)/r=\omega\,j_{\ell=0}(\omega r)$.

The impossibility of defining radial ingoing and outgoing particles is not 
restricted to the Minkowski space-time, this holds also for arbitrary 
spherically symmetric metrics
\begin{eqnarray}
ds^2=g_{00}(r)\,dt^2+g_{11}(r)\,dr^2+r^2d\Omega^2
\,,
\end{eqnarray}
provided that the coefficients of the metric $g_{00}$ and $g_{11}$
are smooth and analytic functions.
Such functions can be Taylor expanded
\begin{eqnarray}
g_{00}(r)=g_{00}(0)+g_{00}''(0)\,\frac{r^2}{2}+{\cal O}(r^3)
\,,
\end{eqnarray}
where $g_{00}'(0)$ and $g_{11}'(0)$ have to vanish for smoothness.
After the separation of the angular variables with 
${\cal Y}_{\ell m}(\vartheta,\varphi)$ the radial dependence of the 
eigenfunctions is governed by a second-order ordinary differential equation 
in $r$.
Provided its coefficients are smooth and regular the solutions of such an 
equation are uniquely determined by the first two (non-vanishing)
terms of their Laurent expansion.
For the evaluation of these initial data only the terms $g_{00}(0)$, 
$g_{00}'(0)=0$, $g_{11}(0)$  and $g_{11}'(0)=0$ are of relevance. 
For that purpose the radial part of the metric can be approximated by 
$ds^2=g_{00}(0)\,dt^2+g_{11}(0)\,dr^2$.
Ergo the behaviour of the corresponding eigenfunctions is 
(up to a simple scale transformation with $g_{00}(0)$ and $g_{11}(0)$
respectively)
asymptotically ($r\rightarrow0$) the same as in the Minkowski space-time.  
Consequently, also in these more general spherically symmetric metrics there 
exists exactly {\em one} eigenfunction for given $\omega,\ell,m$ which 
forbids the definition of radial ingoing and outgoing particles. 

In view of conservation law arguments the nonexistence of ingoing and
outgoing particles in regular space-times appears very plausible: 
Every ingoing component will bounce off at the origin after some
period of time and eventually turn into outgoing and vice versa.  
  
\subsection{Rindler metric}\label{rindler}

As stated in Section \ref{precond} the particle interpretation
crucially depends on the selection of a particular time-like Killing
vector. In the following we shall consider an example where this
dependence will become more evident. In the previous treatments we
focused on the Killing vector mediating the Minkowski time translation
symmetry. Of course this Killing field corresponds to usual observers at
rest. But there exist further time-like Killing vectors in the
Minkowski space-time -- associated with special Lorentz boosts --
which result in a deviating particle interpretation.    
 
Starting with the 1+1 dimensional Minkowski metric $ds^2=dt^2-dx^2$  
and performing the coordinate transformation
\begin{eqnarray}
\label{rindlerkotrafo}
t &=& \rho\sinh\kappa\tau
\,,
\nonumber\\
x &=& \rho\cosh\kappa\tau
\,,
\end{eqnarray}
one arrives at the Rindler metric $ds^2=\kappa^2\rho^2d\tau^2-d\rho^2$.
The quantity $\kappa$ is called the surface gravity, see e.g. \cite{townsend}.
For fixed $\rho$ the transformation describes an accelerated motion.
With respect to the new time coordinate $\tau$ the Rindler metric is static 
and thus allows for a particle definition according to Section \ref{pdef}. 
These particles may be interpreted as those seen by an accelerated observer. 
The corresponding $\cal K$-operator can be cast into the form
\begin{eqnarray}
{\cal K}=-\kappa^2\,
\rho\,\frac{\partial}{\partial\rho}\,\rho\,\frac{\partial}{\partial\rho}
\,,
\end{eqnarray}
with $G=(0<\rho<\infty)$.
The surface term (see Sec. \ref{eom}) at $\rho=0$ vanishes without imposing 
any condition on the field $\Phi$ due to $\sqrt{-g}=\kappa\rho=0$ at $\rho=0$.
Indeed, the Rindler metric possesses a horizon there.
Since the occurrence of this horizon depends on the choice of the
coordinates and thereby on the observer, it is a particle horizon 
(with respect to all world-lines $\rho={\rm const}$) but not an event
(or apparent, etc.) horizon, see Sec. \ref{precond}
and \cite{visser,ellis}.  

For further investigations it is convenient to introduce the tortoise 
coordinate $\rho_*=\ln(\kappa\rho)/\kappa$.
In terms of this coordinate the metric reads
$ds^2=e^{2\kappa\rho_*}\left(d\tau^2-d\rho^2_*\right)$
resulting in the operator 
\begin{eqnarray}
{\cal K}=-
\frac{\partial^2}{\partial\rho^2_*}
\,,
\end{eqnarray}
with $G=(-\infty<\rho_*<\infty)$.
As a consequence, the spectrum is twice degenerated and the corresponding 
eigenfunctions read $\sin(\omega\rho_*)$ and $\cos(\omega\rho_*)$ or 
$\exp(\pm i\omega\rho_*)$, respectively.  
Returning to the coordinate $\rho$ the eigenfunctions behave as
$\exp(\pm i\omega\ln(\kappa\rho)/\kappa)=(\kappa\rho)^{\pm i\omega/\kappa}$.
Even though the domain is bounded $G=(0<\rho<\infty)$ there are two 
eigenfunctions per eigenvalue which allows for the definition of left-moving  
and right-moving particles.
This indicates the absence of real boundary conditions on the field $\Phi$ at 
the horizon $\rho=0$.
In this regard the horizon is the opposite of a mirror.
Even for a finite domain $G=(0<\rho<L)$ the spectrum of the $\cal K$-operator 
is still continuous due to the horizon: $G=(-\infty<\rho_*<L_*)$. 

\subsection{Unruh effect}\label{unruh}

After having performed a particle definition for the Minkowski and the 
Rindler observer, the question about the relationship of these two approaches 
arises.
Evaluating the expectation value of the number of Rindler particles in the 
Minkowski vacuum one obtains a thermal distribution function, a consequence 
of the Unruh \cite{unruh} effect.
This effect demonstrates manifestly that different observers may obey   
distinct particle interpretations. In consequence the vacuum may
depend on the particular Killing vector.

One way (see also \cite{wipf}) to calculate the expectation values
explicitly is based on the Bogoliubov coefficients
\begin{eqnarray}
\label{gammaB}
\beta_{\Gamma\Lambda}=
i\int d\Sigma^\mu\;F_\Gamma^{\rm M}\;
\stackrel{\leftrightarrow}{\partial}_\mu\;
F_\Lambda^{\rm R}
\,.
\end{eqnarray}
The generalised Minkowski eigenfunctions are labelled by $\Gamma=(\xi,\omega)$
\begin{eqnarray}
F_\Gamma^{\rm M}(\underline x)=F_{(\xi,\omega)}^{\rm M}(x,t)
={\cal N}_{\rm M}\frac{\exp(-i\omega t)}{\sqrt{\omega}}
\;e^{i\xi\omega x}
\,,
\end{eqnarray}
where $\xi=\pm1$ distinguishes the left-moving and right-moving particles.
(${\cal N}_{\rm M}$ denotes a normalisation factor.) 
In analogy the generalised Rindler eigenfunctions read
\begin{eqnarray}
F_\Lambda^{\rm R}(\underline x)=F_{(\xi',\omega')}^{\rm R}(\rho,\tau)=
{\cal N}_{\rm R}
\frac{\exp(-i\omega'\tau)}{\sqrt{\omega'}}
(\kappa\rho)^{i\xi'\omega'/\kappa} 
\,.
\end{eqnarray}
With the choice for the surface $\Sigma=\{\tau=0,0<\rho<\infty\}$ the surface 
element takes the form $d\Sigma^0=g^{00}\sqrt{-g}\,d\rho=d\rho/(\kappa\rho)$.
At this surface the Minkowski coordinates are simply given by $t=0$, $x=\rho$ 
and the derivative transforms according to 
$\partial_\tau=\kappa\rho\,\partial_t$.
Putting all this together, the $\beta$-coefficient transforms into
\begin{eqnarray}
\label{gammaF}
\beta_{(\xi,\omega);(\xi',\omega')}=
{\cal N}_{\rm MR}
\int\limits_0^\infty\frac{d\rho}{\kappa\rho}\;
\frac{\omega'-\kappa\rho\,\omega}{\sqrt{\omega\omega'}}\;
e^{i\xi\omega\rho}\,(\kappa\rho)^{i\xi'\omega'/\kappa}
\,.
\end{eqnarray} 
This integral involves generalised eigenfunctions 
(corresponding to $\hat A_\Gamma$) and has to be understood in a 
distributional sense.
For well-defined expressions (such as $\hat a_I$) we have to insert a 
convergence factor, for instance 
$(\kappa\rho)^\varepsilon\,\exp(-\varepsilon\kappa\rho)$.
The existence of the limit $\varepsilon\downarrow0$ confirms the possibility 
of approximating the singular eigenfunctions by regular quantities.
After this procedure we may make use of the formula \cite{abra}
\begin{eqnarray}
\label{gamma}
\int\limits_0^\infty du\;e^{-uw}\;u^{z-1}=w^{-z}\,\Gamma(z)
\,,
\end{eqnarray}
which holds for $\Re(w)>0$ and $\Re(z)>0$, and we arrive at
\begin{eqnarray}
\label{gammabeta}
\beta_{(\xi,\omega);(\xi',\omega')}=
{\cal N}_{\rm MR}
\frac{1+\xi'\xi}{\kappa}\;
\sqrt{\frac{\omega'}{\omega}}\;\Gamma(i\xi'\omega'/\kappa)\;
\left(-i\xi\omega/\kappa+\varepsilon\right)^{-i\xi'\omega'/\kappa}
\,.
\end{eqnarray}
Calculating the remaining Bogoliubov coefficient
$\alpha_{(\xi,\omega);(\xi',\omega')}$ one gets nearly the same
expression but with a positive sign in front of the term
$i\xi\omega/\kappa$. 
Therefore both coefficients merely contribute for particles moving in
the same "direction"
$\beta_{(\xi,\omega);(\xi',\omega')}\sim\delta_{\xi,\xi'}$, 
respectively, 
$\alpha_{(\xi,\omega);(\xi',\omega')}\sim\delta_{\xi,\xi'}$.
Now it is possible to compare both quantities.
As said before, the only difference between $\alpha$ and $\beta$ is
the sign in front of the term $\omega/\kappa$.
Dividing the two coefficients all other terms cancel and the convergence 
factor $\varepsilon$ determines the side of the branch cut of the logarithm 
in the complex plane. Hence we find 
\begin{eqnarray}
\label{bexpa}
\beta_{(\xi',\omega);(\xi',\omega')}=
\exp(-\pi\omega'/\kappa)\,
\alpha_{(\xi',\omega);(\xi',\omega')}
\,.
\end{eqnarray}
An alternative way to obtain this important result is based on the
analytic continuation into the complex plane. 
For that purpose we define slightly modified Bogoliubov coefficients via
\begin{eqnarray}
\beta^{\rm c}_{\xi,\xi'}(\omega,\omega')
=
\sqrt{\omega\omega'}\beta_{(\xi,\omega);(\xi',\omega')}
\,,
\end{eqnarray}
and in analogy the $\alpha$-coefficient.
In view of Eqs. (\ref{gammaB})--(\ref{gammaF}) the modified Bogoliubov 
coefficients can be analytically  continued to the complex
$\omega$-plane where the relation 
\begin{eqnarray}
\alpha^{\rm c}_{\xi,\xi'}(\omega,\omega')
=
\beta^{\rm c}_{\xi,\xi'}(-\omega,\omega')
\end{eqnarray}
holds.
Inserting this equality into Eq. (\ref{gammabeta}) reproduces
Eq. (\ref{bexpa}).
In order to evaluate the absolute value squared of the
$\beta$-coefficient  we may utilise the identity \cite{abra} 
\begin{eqnarray}
\Gamma(z)\Gamma(-z)=-\frac{\pi}{z\sin\pi z}
\end{eqnarray}
to obtain the final result
\begin{eqnarray}
\label{beta^2}
\left|\beta_{(\xi,\omega);(\xi',\omega')}\right|^2=
\frac{8\pi{\cal N}_{\rm MR}^2}{\kappa\omega}\,
\frac{\delta_{\xi,\xi'}}{\exp(2\pi\omega'/\kappa)-1}
\,.
\end{eqnarray}
In view of the remaining $\omega$-integration the number of Rindler particles 
in the Minkowski vacuum diverges.
This result can also be re-derived using the well-known unitarity relation
\begin{eqnarray}
\label{unit}
\sumint{\Gamma}\alpha_{\Lambda\Gamma}\alpha_{\Upsilon\Gamma}^*-
\beta_{\Lambda\Gamma}\beta_{\Upsilon\Gamma}^*=
\delta(\Lambda,\Upsilon)
\,,
\end{eqnarray}
where $\Gamma$ symbolises the Minkowski index.
This equality reflects the completeness of the Minkowski solutions. 
Special care is required concerning the derivation of an analogue expression 
involving the Rindler functions since these solutions are restricted to the 
Rindler wedge and thereby they are not complete in the full Minkowski 
space-time.

Inserting Eq. (\ref{bexpa}) and considering the singular coincidence
$\Lambda=\Upsilon$ it follows
\begin{eqnarray}
\label{Nunruh}
N_{(\xi',\omega')}
&=&
\bra{0_{\rm M}}\hat N_{(\xi',\omega')}^{\rm R}\ket{0_{\rm M}}=
\sumint{(\xi,\omega)}
\left|\beta_{(\xi,\omega);(\xi',\omega')}\right|^2
\nonumber\\
&=&
\frac{\delta(\omega,\omega)}{\exp(2\pi\omega'/\kappa)-1}=
\frac{{\cal N}_V V}{\exp(2\pi\omega'/\kappa)-1}
\,.
\end{eqnarray}
In the last step we have used the results of Section \ref{continuum}.
Recalling the argumentation made there we come to the conclusion that the 
divergence of $N_{\Gamma}$ is {\em necessary} for a thermal behaviour. 

The same calculation can be performed with well-defined operators 
$\hat n_I$ corresponding to localised wave packets. 
For an appropriately chosen basis $e_I(\f{r})$ the coefficients
$\beta_{IJ}$ are up to normalisation factors approximately the same as 
the $\beta_{\Gamma\Lambda}$ evaluated above.
But in this case the results for 
$\left<\hat n_I\right>=\sum_J|\beta_{IJ}|^2$ are finite
owing to $\delta_{II}=1$.
Another explanation is the fact, that the $\beta_{IJ}$
for arbitrary frequencies $\omega$ do not coincide with
the $\beta_{\Gamma\Lambda}$ (due to the localised character
of the wave packets) which makes the $\omega_I$-summation finite.
One way to perform technically such a calculation involving localised 
quantities is to insert a convergence factor with a finite $\varepsilon$
similar to the comment after Eq. (\ref{gammaF}).
Accordingly this finite $\varepsilon$ enters the $\beta$-coefficients 
and causes a finite result of the $\omega_I$-summation and thereby  
a finite number of created particles as well.
Omitting the corresponding normalisation factor the infinite volume 
divergence can be restored in the limit $\varepsilon\downarrow0$.

\subsection{KMS condition}\label{kms}

From a strictly axiomatic point of view the divergent result in
Eq. (\ref{Nunruh}) in the last section may not be completely convincing.
However, it is possible to show more rigorously that the Minkowski
vacuum indeed behaves as a thermal state when analysed by a Rindler observer.
This can be done by employing the Kubo-Martin-Schwinger (KMS)
condition \cite{kubo,martin+schwinger}. 
A KMS state $\langle\cdot\rangle_T$ is defined as a  
time-translationally invariant state which satisfies the following condition
\begin{eqnarray}
\langle\hat U(t)\hat V(t')\rangle_T
=
\langle\hat V(t')\hat U(t+i/T)\rangle_T
\end{eqnarray}
for all observables $\hat U$ and $\hat V$ and some temperature $T$.
It can be shown that if the (irreducible) algebra of observables
possesses  a well-defined matrix-representation then the KMS state
corresponds to the usual canonical ensemble
\begin{eqnarray}
\langle\hat U\rangle_T
=
{\rm Tr}\left\{\hat U\,\frac{\exp(-\hat H/T)}{Z}\right\}
\,.
\end{eqnarray}
One might wonder at the fact that the Minkowski vacuum, i.e. a pure
state, displays thermal features -- usually connected with mixed states.
This can be explained by the thermo-field formalism, see
e.g. \cite{thermo-field}. I.e., a pure state of a quantum system
transforms into a mixed state after averaging over a subsystem
owing to the correlations between the different subsystems.
As a result of the particle horizon at $\rho=0$ the Rindler observer
is causally separated from a part of the Minkowski space-time and
does therefore indeed regard the Minkowski vacuum as a mixed state. 

To show that the Minkowski vacuum displays the temperature
$T=\kappa/(2\pi)$ we consider the corresponding two-point Wightman
\cite{wightman} function.
The Wightman axioms (in particular the spectral condition) imply that
this bi-distribution can be considered as the boundary value of an
analytic function.
Hence we may restrict to the space-like region for reasons of
simplicity where the two-point function assumes the form
\begin{eqnarray}
W(\underline x,\underline x')
=
\langle\hat\Phi(\underline x)\hat\Phi(\underline x')\rangle
=
-\frac{(2\pi)^{-2}}{(\underline x-\underline x')^2}
\end{eqnarray}
for 3+1 dimensions. 
In 1+1 dimensions it behaves as $\ln[(\underline x-\underline x')^2]$ 
which does not alter the following considerations.
Since for a free field all $n$-point functions can be derived from
this 2-point function it contains all information about the theory.

Now we may consider the two-point function in terms of Rindler
coordinates. The $t,x$-contribution to the geodesic distance
transforms according to Eq. (\ref{rindlerkotrafo}) into
\begin{eqnarray}
(t-t'\,)^2-(x-x'\,)^2=2\rho\,\rho'\,\cosh\left(\kappa[\tau-\tau'\,]\right)
-\rho^2-{\rho'\,}^2
\,.
\end{eqnarray}
As a result the two-point function is periodic along the imaginary
Rindler time axis and thus satisfies the KMS condition for the
temperature $T=\kappa/(2\pi)$.
This result confirms the considerations in the previous section and
justifies the identification of the UV-divergence occurring there with
the infinite volume divergence of the Rindler space.

It can be shown quite generally that only the KMS state corresponding
to the temperature $T=\kappa/(2\pi)$ satisfies the Hadamard
\cite{hadamard} condition (local stability) in the complete Rindler
space-time (and in particular at the horizon), see \cite{haag}.
The Hadamard requirement demands the singularity of the two-point 
function ($1/s^2$ and $\ln s^2$) to be independent of the state,
i.e. it is only determined by the structure of the space-time, see
also \cite{kay+wald,verch,fulling-buch,wipf}.
This property ensures the validity of the point-splitting
renormalisation technique, cf. \cite{wald-t}.
As it will become more evident in Section \ref{ext}, an analogue idea
can be employed to derive the Hawking effect.

\section{Black holes}\label{schwarz}

In this Section we are going to apply the formalism presented in Sec. 
\ref{pdef} to one of the most fascinating curved space-time structures, 
the black hole.
Various coordinate systems which represent this object are known. 
For our purpose we have to demand a static metric with a time coordinate $t$ 
corresponding to a Killing vector.
Because the black hole space-time becomes asymptotically flat, another 
requirement is the coincidence of this time coordinate $t$ with the usual 
Minkowski time of an observer at spatial infinity.
All this requisites are fulfilled by the Schwarzschild coordinates 
$t,r,\vartheta,\varphi$ for which the black hole metric reads
\begin{eqnarray}
ds^2=h(r)\,dt^2-\frac{dr^2}{h(r)}
-r^2d\vartheta^2-r^2\sin^2\vartheta\,d\varphi^2
\,.
\end{eqnarray}
Other coordinates, e.g. Kruskal, Eddington-Finkelstein, etc. are not suitable 
for the above reasons.    
As the Schwarzschild coordinates measure time and length scales with respect
to an observer at fixed spatial distance to the black hole all results
obtained later refer to this observer.

In order to describe a (non-extreme) black hole with a horizon at
$r=R$ and a surface gravity $\kappa>0$ the function $h$ obeys the properties 
(see e.g. \cite{townsend})
\begin{eqnarray}
h(R)=0\quad,\quad\kappa=\frac{1}{2}\,h'(R)
\end{eqnarray}
and also $h(r>R)>0$ together with $h(r\rightarrow\infty)=1$.
With the aid of this function $h$ it is possible to consider the rather 
general case of a static black hole, for example the Schwarzschild metric 
with $h=1-R/r$. 

Using these coordinates the canonical conjugate momenta turn out to be 
$\hat\Pi=\partial_t\hat\Phi/h$.  
In terms of these momenta the formal expression for the Hamiltonian density 
can be cast into the following form
\begin{eqnarray}
\hat{\cal H}=\frac{h}{2}\,\hat\Pi^2+
\frac{h}{2}\,\left(\partial_r\hat\Phi\right)^2+
\frac{1}{2r^2}\left(\partial_\vartheta\hat\Phi\right)^2+
\frac{1}{2r^2\sin^2\vartheta}\left(\partial_\varphi\hat\Phi\right)^2
\,.
\end{eqnarray}
The fields $\hat\Phi(\f{r},t)$ as well as their momenta $\hat\Pi(\f{r},t)$ are 
operator-valued distributions (see also Section \ref{pdef}).
Consequently, the Hamiltonian density above is not well-defined.
In analogy to Sec. \ref{canon} it may only be considered as a formal
expression until an appropriate regularisation method, for instance
the point-splitting technique (see e.g. \cite{wald-t}), has been applied.  

It is possible to split up the Hamiltonian $\hat H$ of the field $\hat\Phi$ 
into two parts $\hat H=\hat H_>+\hat H_<$ that account for the 
interior $\hat H_<$ and the exterior $\hat H_>$ region of the 
black hole, respectively
\begin{eqnarray}
\hat H_>&=&\int d^3\f{r}\;\hat{\cal H}\;\Theta(r-R)
\,,
\nonumber\\
\hat H_<&=&\int d^3\f{r}\;\hat{\cal H}\;\Theta(R-r)
\end{eqnarray}
with the Heaviside step function $\Theta$ and the volume element  
$d^3\f{r}=\sqrt{-g}\,d^3x=r^2\sin\vartheta\,dr\,d\vartheta\,d\varphi$.
Employing the equal time commutation relations 
\begin{eqnarray}
[\hat\Phi(\f{r},t),\hat\Phi(\f{r}',t)]=
[\hat\Pi(\f{r},t),\hat\Pi(\f{r}',t)]=0
\quad;\quad
[\hat\Phi(\f{r},t),\hat\Pi(\f{r}',t)]=
i\delta^3(\f{r}-\f{r}')
\,,
\end{eqnarray}
where $t$ denotes the Schwarzschild time and represents a Killing vector,
one observes that the two parts of the Hamiltonian commute 
\begin{eqnarray}
\left[\hat H_>,\hat H_<\right]=0
\,.
\end{eqnarray}
In the language of point-splitting,
cf. \cite{wald-t} and the remarks in Section \ref{kms}, 
the divergent terms of the Hamiltonian density are independent of the state and 
therefore pure $\mathbb C$-numbers which do not contribute to the commutator.
The remaining (convergent) operator-valued components commute because of 
$h(r=R)=0$.
The same result can be obtained by means of normal ordering or the
regularisation described in Eq. (\ref{regHam}). Due to $h(r=R)=0$ the
$\cal K$-operator and the operators projecting onto the interior,
respectively, exterior domain commute. Hence it is possible to select a basis
$b_I$ for the inside and outside region separately such that the
Hamiltonian possesses no mixing terms.  

Accordingly, the separation $\hat H=\hat H_>+\hat H_<$ represents 
two independent systems.
This fact displays one advantage of the Schwarzschild coordinates because 
there {\em is} a horizon at $r=R$.
The consistency with the results of Section \ref{precond} can be demonstrated 
if one considers the spatial surface term
$dS_i\;g^{ij}=d^2x\;\sqrt{-g}\;n_i\;g^{ij}=
d\vartheta\;d\varphi\;r^2\sin\vartheta\;g^{rr}$
which indeed vanishes for $r=R$.
As a consequence, it is impossible to transport matter (energy or information)
across the horizon, nothing can come out or fall into the black hole.  
Of course, this holds only for a fixed metric, i.e. if one neglects the 
back-reaction.
Without this restriction it is possible that the horizon increases due to 
the in-falling matter, and swallows it.
It should be emphasised again that all of our assertions refer to an
observer at a fixed spatial distance to the black hole and therefore
not necessarily to a free falling one.
 
\subsection{Black hole exterior}\label{ext}

In the following we restrict our considerations to the domain outside the 
black hole governed by $\hat H_>$.
The properties of the interior will be discussed in the next Section.
The exterior region $G=\{r>R\}$ fulfils the conditions imposed in 
Sec. \ref{precond} which allows for a particle definition.
As a result, the $\hat H_>$-part of the Hamiltonian can be diagonalised 
formally via 
\begin{eqnarray}
\hat H_>=\sumint{\Gamma}\omega_\Gamma\hat N_\Gamma^{\rm BH}+
E_\infty
\,,
\end{eqnarray}
where $E_\infty$ denotes the divergent zero-point energy.

In order to isolate the features that are specific for black holes,
the most interesting region is the neighbourhood of the horizon 
$r\approx R$. 
To investigate the behaviour in this zone we introduce a dimensionless 
variable $\chi$ with
\begin{eqnarray}
\chi=2\kappa(r-R)\rightarrow
h=\chi\left(1+{\cal O}(\chi)\right)
\,.
\end{eqnarray}
Without loosing the static character of the metric it is possible to perform 
a radial coordinate transformation for $r>R$ via
\begin{eqnarray}
r_*=\int\frac{dr}{h}=
\frac{\ln\chi}{2\kappa}+{\cal O}(\chi)
\,.
\end{eqnarray}
The new radial $r_*$ coordinate is called the the Regge-Wheeler
tortoise coordinate. 
According to the above arguments it is sufficient to cover the region outside
the horizon by the new coordinate.
The function $h$ and the original radial variable $r$ have to be considered 
as functions of the introduced coordinate: $r=r(r_*)$ and
$h=h(r_*)=h(r[r_*])$.
The tortoise coordinate has the advantage of a very simple form of the 
$\cal K$-operator 
\begin{eqnarray}
{\cal K}=-\frac{1}{r^2}\frac{\partial}{\partial r_*} r^2  
\frac{\partial}{\partial r_*}-h\nabla^2_{\vartheta\varphi}
=-\frac{\partial^2}{\partial r_*^2}+{\cal O}(\chi)
\,,
\end{eqnarray}
together with $G=\{r_*\in{\mathbb R}\}$.
The generalised eigenfunctions $F_\Gamma^{\rm BH}(\underline x)$ 
of this operator 
behave ${\cal O}(\chi)$-approximately as $\exp(\pm i\omega r_*)$ and 
after the separation of the angular variables they can be written as follows
\begin{eqnarray}
\label{eigenBH}
F_\Gamma^{\rm BH}(\underline x)=
F_{\xi\omega\ell m}^{\rm BH}(t,\chi,\vartheta,\varphi)
={\cal N}^{\rm BH}_{\omega\ell}\,
\frac{e^{-i\omega t}}{\sqrt{\omega}}\,
\chi^{i\xi\omega/(2\kappa)}\,{\cal Y}_{\ell m}(\vartheta,\varphi)
\left(1+{\cal O}(\chi)\right)
\,.
\end{eqnarray}
${\cal N}^{\rm BH}_{\omega\ell}$ symbolises a normalisation factor which may 
without any loss of generality chosen to be independent of $\xi$.
These eigenfunctions are rapidly oscillating near the horizon.

By inspection, we recognise the occurrence of two generalised
eigenfunctions for a given frequency $\omega$ and fixed 
angular dependence $\ell,m$ distinguished by $\xi=\pm1$.  
Thus the definition of ingoing and outgoing particles is possible 
in this case.
(It should be mentioned that potential scattering effects cause  
slight deviations from the purely ingoing and outgoing behaviour 
in Eq. (\ref{eigenBH}) at $r\rightarrow\infty$.
However, this way of definition does not alter the conclusions.)
This -- perhaps surprising -- fact can be elucidated in the
following way.
The horizon separates the space into two independent
domains (interior and exterior) and prevents the 
field modes outside from being influenced by the
effective "boundary condition" at $r=0$.
In view of the study of the $\cal K$-operator in terms of the
tortoise coordinate $r_*$ one may consider the horizon as
some new kind of spatial infinity  ($r_*\rightarrow-\infty$)
in addition to $r\rightarrow\infty$.

Also for the black hole example the horizon acts opposite to a mirror,
cf. Section \ref{rindler}.
Even for the scenario of a black hole which is enclosed in a large box 
with Dirichlet boundary conditions the spectrum of the operator
$\cal K$ is still continuous -- but now not degenerated.    

For a black hole in an asymptotically flat (unbounded) space-time there 
are two contributions to the infinite volume divergence 
(see Sec. \ref{continuum}) $\delta(\Gamma,\Gamma)={\cal N}_VV$:
firstly, the usual infinity $r,r_*\rightarrow\infty$  and secondly,
the effective infinity at the horizon $r\rightarrow R$
respectively $r_*\rightarrow-\infty$.
The former divergence $\delta_+(\Gamma,\Gamma)$ does also arise in the 
(unbounded) Minkowski space-time -- but not inside a finite box
(e.g. with Dirichlet boundary conditions) -- whereas the latter
divergence $\delta_-(\Gamma,\Gamma)$ is restricted to the scenario of
a black hole, but it is not affected by a finite box.
E.g., the expectation value of the number "operator" 
$\hat N_\Gamma$ in any  KMS state (with a non-vanishing temperature)
contains the complete divergence $\langle\hat N_\Gamma\rangle_T\sim
\delta_+(\Gamma,\Gamma)+\delta_-(\Gamma,\Gamma)$.
One important example is the Israel-Hartle-Hawking
\cite{israel,hartle} state, the KMS state corresponding to the Hawking
temperature $T=\kappa/(2\pi)$.  
For large radial distances to the black hole the (renormalised)
expectation value of the energy-momentum tensor evaluated in this state
approaches a constant value (proportional to $T^4$).
In contrast, for the Unruh \cite{unruh} state -- the state describing 
the black hole evaporation -- the (renormalised) energy density
decreases with $1/r^2$ for large $r$.
As a consequence, the expectation value of the number of particles in
this state does not display the complete divergence $\delta_+(\Gamma,\Gamma)$.

It might be interesting to illustrate the point above with the aid of
the Bogoliubov coefficients:
If we consider the spherically symmetric collapse of a star to a black hole 
the metric outside the initial radius of the star does not change 
(Birkhoff theorem).
Ergo the behaviour of the modes at very large radial distances $r$ is not 
affected by the collapse.
Accordingly, this region does not contribute to the
$\beta_{\omega,\omega'}$-coefficients and generates a
$\delta(\omega-\omega')$-term for the $\alpha_{\omega,\omega'}$-coefficients,
cf. also \cite{hawking}.
Recalling the unitarity relation for the Bogoliubov coefficients in
Eq. (\ref{unit}) we arrive at the conclusion that exactly this term
generates the $\delta_+(\Gamma,\Gamma)$-part of the infinite volume
divergence.
Following Hawking \cite{hawking} we assume that 
-- for large initial frequencies $\omega$ --
the Bogoliubov coefficients are related via Eq. (\ref{bexpa}) in analogy 
to Section \ref{unruh}.
Proceeding in the same way as in that Section we observe that 
the $\omega$-integration of the absolute value squared of 
the $\beta_{\omega,\omega'}$-coefficients is UV-divergent again.
But in contrast to the Unruh effect this divergence does not contain
$\delta_+(\Gamma,\Gamma)$, but only $\delta_-(\Gamma,\Gamma)$
owing to the unitarity relation (\ref{unit}).
Consequently the Minkowski vacuum is a KMS state with respect to the
Rindler observer, but it does not transform into a KMS state during
the collapse to a black hole 
(if we assume the space-time to be asymptotically flat and 
therefore unbounded).  
Hawking derived the relation (\ref{bexpa}) only for the finally outgoing
particles. But even if this relation would hold for both, 
the (finally) ingoing and outgoing particles, the state would still
contain less particles than the corresponding KMS state. 

As it became evident in the previous considerations, the vicinity of
the horizon of a black hole displays many similarities to the scenario
of the Unruh effect in Section \ref{unruh}. 
Indeed, with $\chi=\kappa^2\rho^2$ the black hole metric approaches
the Rindler metric in that region 
\begin{eqnarray}
\label{schwarzrindler}
ds^2=\left(\kappa^2\rho^2dt^2-d\rho^2-R^2d\Omega^2\right)
\left(1+{\cal O}(\chi)\right)\,,
\end{eqnarray}
together with the angular part $d\Omega^2$.
This observation motivates an argumentation analogue to that at the
end of Sec. \ref{kms}, cf. also \cite{deser}. 
Indeed, it is possible to prove \cite{fredenhagen} that for a state 
fulfilling the Hadamard requirement 
(among other conditions, see \cite{fredenhagen})
throughout the complete space-time (and in particular at the horizon)
the asymptotic expectation values correspond to the Hawking temperature.
The ground state of the quantum field (the Boulware state)
as well as every KMS state (with an arbitrary temperature) obey the
Hadamard singularity structure {\em away} from the horizon, 
see \cite{verch}.
But only that KMS state that corresponds to the Hawking temperature
$T=\kappa/(2\pi)$ -- i.e. the Israel-Hartle-Hawking state -- matches
the Hadamard condition {\em at} the horizon. 
(The same holds true for the Unruh state.)
It can be shown that the Hadamard condition is conserved during the
dynamics of a $C^\infty$ space-time.
Accordingly, if the collapse of a star to a black hole can be
described by a $C^\infty$-metric, the consideration above can be used
to deduce the Hawking effect.
(The Minkowski vacuum is of course also a Hadamard state.)
However, dropping the assumption of a $C^\infty$ space-time
the situation becomes less clear.

If we compare the outcome of this Section with the Minkowski example, 
we arrive at the conclusion that the formation of the horizon causes a 
bifurcation in a double sense:

The total Hamiltonian of the field $\hat H$ splits up into two commuting 
parts $\hat H_<$ and $\hat H_>$ which account for two independent 
(physical complete) regions $r<R$ and $r>R$, respectively.

Before the horizon has been formed there exists only one generalised 
eigenfunction for every given frequency and fixed angular behaviour. 
This property forbids the definition of ingoing and outgoing particles 
(see Section \ref{ingoing}). 
After the horizon has been formed the spectrum is twice degenerated and the 
definition of ingoing and outgoing particles becomes possible.

\subsection{Black hole interior}\label{int}

Our previous investigations focused on the exterior of the black hole.
As indicated before we shall now take the interior region into account. 
Inside the (non-extreme) black hole it yields $h(r)<0$ and therefore
$g_{tt}<0$, $g_{rr}>0$, $g_{\vartheta\vartheta}<0$ and
$g_{\varphi\varphi}<0$.
As a consequence the signature of the metric is degenerated and thus the 
particle and vacuum definition proposed in Section \ref{pdef} does not apply.
However, it is still possible to obtain a self-adjoint
$\cal K$-operator governing the dynamics of the system. 
But for this purpose some modifications are necessary with the result 
that $\cal K$ is {\em not} given by ${\cal D}^\dagger\overline{\cal D}$ 
and therefore {\em not} non-negative.
As we shall see later the negative parts of $\cal K$ correspond to
unstable fields modes. 

At first the scalar product of the interior region 
$\scalar{\cdot}{\cdot}_1^<$ has to be defined
with $|g^{00}|$ instead of $g^{00}$ in Eq. (\ref{scalarproduct}) in
order to obtain a positive-definite bilinear form.
For reasons of simplicity we restrict our further considerations to
the Schwarzschild metric $h(r)=1-R/r$ and start with the functions 
\begin{eqnarray}
C^\infty_0(G_<)={\rm lin}
\left\{
C^\infty_0(0<r<R)
\otimes
C^\infty({\mathbb S}_2)
\right\}
\,,
\end{eqnarray}
where ${\mathbb S}_2$ denotes the 2-sphere of $\vartheta$ and
$\varphi$.
Again the Hilbert space $L_2(G_<,|g^{00}|)$ is given by the completion 
of all these functions with respect to the (redefined) scalar product.

The degenerated signature permits the definition of a scalar product
containing $g^{ij}$.
Accordingly, the subsequent steps in Sec. \ref{pdef} cannot be adopted
here.
In particular we cannot introduce an operator $\cal D$ such that the
self-adjoint $\cal K$-operator is represented by the absolute value
squared  of $\cal D$.   
Instead we may define an operator ${\cal K}_0$ via
\begin{eqnarray}
{\cal K}_0\,:\,C^\infty_0(G_<)\,\subset\,L_2(G_<,|g^{00}|)
& \rightarrow & L_2(G_<,|g^{00}|) \nonumber\\
\phi & \rightarrow &
-\frac{h}{r^2}\frac{\partial}{\partial r}hr^2  
\frac{\partial\phi}{\partial r}-h\nabla^2_{\vartheta\varphi}\phi
\,.
\end{eqnarray}
The second term at the r.h.s. of the above expression for ${\cal K}_0$
generates the negative parts of this operator.
These negative parts originate from the angular derivatives and cannot 
be obtained in a purely radial symmetric consideration.

Obviously ${\cal K}_0$ is Hermitian with respect to the scalar
product containing the weight $|1/h|$ (and $\sqrt{-g}=r^2\,\sin\vartheta$).
In addition -- since ${\rm Def}({\cal K}_0)=C^\infty_0(G_<)$ is dense
in the underlying Hilbert space $L_2(G_<,|g^{00}|)$ -- it is densely
defined and therefore symmetric.

Now we can make use of the following theorem (see e.g. \cite{weidmann}):
Every symmetric and $\mathbb C$-real operator acting on a complex
Hilbert space possesses (at least one) self-adjoint extension(s). 
As a result we will always find a self-adjoint operator $\cal K$ 
(as an appropriate extension of ${\cal K}_0$) governing the
dynamics of the field. 
In terms of $\cal K$ the Lagrange function in Eq. (\ref{lagGij})
for the interior domain assumes the simple form
\begin{eqnarray}
L_<=
-\frac{1}{2}\scalar{\dot\Phi}{\dot\Phi}^<_1
+\frac{1}{2}\sca{\Phi}{\cal K}\lar{\Phi}^<_1
\,.
\end{eqnarray}
Note that in contrast to Eq. (\ref{lagK}) the global sign has changed.
However, this global sign does not affect the equation of motion, but
-- as it will become evident later --
the negative parts of the $\cal K$-operator do so.

Since the self-adjoint $\cal K$-operator represents an extension of
the original operator ${\cal K}_0$ these two operators have to
coincide on the subspace $C^\infty_0(G_<)$.
Accordingly, it is possible to construct test functions 
$w(r,\vartheta,\varphi)=w(r)\,{\cal Y}_{\ell m}(\vartheta,\varphi)
\in C^\infty_0(G_<)$ generating negative expectation values of the
${\cal K}$-operator via
\begin{eqnarray}
\sca{w}{\cal K}\lar{w}=\int\limits_0^R dr
\left(
|h|r^2\left|\partial_r w\right|^2
-\ell(\ell+1)\left|w\right|^2
\right)<0
\,.
\end{eqnarray}
If we choose the angular quantum number $\ell$ very large the
expectation value $\sca{w}{\cal K}\lar{w}$ equals negative numbers of
arbitrarily large absolute values, even for normalised test functions
$\scalar{w}{w}=1$.
Hence the spectrum of $\cal K$ is unbounded from below.
(Of course it is also unbounded from above.)
Diagonalising the Hamiltonian by means of a quasi-unitary
transformation $\cal V$ in analogy to Sec. \ref{pdef} yields
\begin{eqnarray}
\hat H_<=-\frac{1}{2}\sumint{\Gamma}
\left(
\hat P_{\Gamma}^2+
\lambda_{\Gamma}\hat Q_{\Gamma}^2
\right)
\,.
\end{eqnarray}
The interior Hamiltonian is still self-adjoint (by Stone's theorem) -- 
but it is {\em not} bounded from above and below.   
Ergo it does not possess a ground state, and a definition of particles 
as excitations over the ground state is impossible. 

As mentioned before, the global sign does not affect the equations of
motion, but the occurring negative eigenvalues 
$\lambda_{\Gamma}$ do so:
The modes $\Gamma$ corresponding to negative eigenvalues 
$\lambda_{\Gamma}$ obey the following equations of motion  
\begin{eqnarray}
\frac{d^2}{dt^2}\hat Q_{\Gamma}=
|\lambda_{\Gamma}|\,\hat Q_{\Gamma}
\,.
\end{eqnarray}
Their solutions $\exp(\pm\sqrt{|\lambda_{\Gamma}|}\,t)$ display a highly 
(linear) unstable behaviour. 

This instability cannot be avoided by introducing an indefinite metric 
of the Fock space \cite{schroer+swieca} if we assume the black hole to 
be formed by a collapse because in this case the Fock space is initially 
well-defined and obeys a positive definite metric
$\braket{\Psi}{\Psi}\geq0$.

One might suspect that the initial conditions are just in such a way
that the exponentially increasing  solutions do not occur, 
cf. also \cite{beholder}. 
Employing an analogue from classical mechanics this situation
corresponds to a point-mass moving on the top of a parabolic hill
which just comes to rest at the zenith of the parabola. 
However, within quantum theory no (regular) stationary state exists in
such a scenario (Heisenberg uncertainty relation).  
Even if the expectation value of the amplitude $\hat Q_{\Gamma}$
vanishes for all times, its variance increases exponentially 
(for late times). 

Since the unstable behaviour described above accounts for the
time-evolution of the (global) modes $\Gamma$ it describes a  
global instability which should not be confused with the concept of
local stability usually associated with the Hadamard condition,
cf. \cite{haag} and \cite{verch}.

It should be mentioned here that potential terms (which we have
omitted in Sec. \ref{eom}) may also give raise to negative parts of
the $\cal K$-operator. E.g., if the assumptions in Section \ref{Kop}
fail and the scalar curvature $\cal R$ assumes negative values over a  
large enough volume the operators $\cal B$ and $\cal K$ are not non-negative.
However, in this situation the $\cal K$-operator is still bounded from 
below (if $\cal R$ does not diverge). Hence only modes up to a
certain quantum number are unstable. These modes are strongly
correlated to the global structure of the space-time.
Special care is required concerning the interpretation of the
instability caused by mass terms. Mass terms that are generated by
the Higgs mechanism occur in the effective Lagrangian for low
excitations and cannot be extrapolated to large amplitudes.    
Restricting ourselves to the massless and minimally coupled scalar
field (as a model for the photon field) only the instability due to
the angular derivatives remains where all these objections do not
apply.  

In order to interpret the instability it might be interesting to
investigate the corresponding proper or generalised eigenfunctions.
Near the horizon (inwards), the modes behave as
\begin{eqnarray}
f_{\Gamma}
\sim
\exp\left(-r_*\,\sqrt{\lambda_{\Gamma}}\right)
\sim
\left(2\kappa[R-r]\right)^{-\sqrt{\lambda_{\Gamma}}/(2\kappa)}
\,.
\end{eqnarray}
Depending on the behaviour at the origin $r=0$ one might expect the
existence of proper eigenfunctions $f_{\Gamma}$
at some points of the negative part of the spectrum.

However, even if no proper and (pointwise/locally defined) generalised
eigenfunctions exist, one may still construct suitable distributions 
$f_{\Gamma}$ with analogous properties \cite{fredenhagen+privat}:
Considering the Schwartz/Sobolev space 
${\mathfrak S}_1(\sigma,{\cal V}) \subset L_2(\sigma,{\cal V})$ 
of all continuous functions over the spectrum $\sigma$ of the 
$\cal K$-operator we may define a Dirac $\delta$-distribution as a
linear functional over this space.
This distribution $\delta_\Gamma=\delta(\lambda,\lambda_\Gamma)$
is then defined within the dual space ${\mathfrak S}_{-1}(\sigma,{\cal V})$.
It represents a generalised eigendistribution of the diagonalised 
$\cal K$-operator ${\cal V}{\cal K}{\cal V}^\dagger\,\delta_\Gamma=
{\cal M}\,\delta_\Gamma=\lambda_{\Gamma}\delta_\Gamma$.
Hence its spatial representation 
$f_\Gamma={\cal V}^\dagger\delta_\Gamma$
exists at least as a distribution over 
${\cal V}^\dagger{\mathfrak S}_1(\sigma,{\cal V})$
(which is dense in $L_2(G_<,|g^{00}|)$)
and describes an eigendistribution of $\cal K$.
The construction described above generates non-vanishing eigendistributions
$f_{\Gamma}$ for all non-singular points (of the spectral measure) $\Gamma$ 
of the spectrum $\sigma$.
Since every open interval of $\sigma$ contains non-singular points 
we can always find an appropriate mode $\Gamma$ where $f_{\Gamma}$ exists.

Using these eigendistributions $f_\Gamma(\f{r})$ we can construct
solutions of the Klein-Fock-Gordon equation of the form
\begin{eqnarray}
\label{unstable}
F_\Gamma(\f{r},t)=
\exp\left(\pm\sqrt{|\lambda_{\Gamma}|}\,t\right)f_\Gamma(\f{r})\,,
\end{eqnarray}
if we choose a mode $\Gamma$ from the negative part of the spectrum.
As a consequence, the equation of motion does not only possess
unstable solutions -- even the degree of the instability 
$\sqrt{|\lambda_{\Gamma}|}$ can be arbitrarily large.
In a vivid description one may speak about an explosion interiorly.

It should be mentioned here that a (partial) negative Hamiltonian, 
i.e. a (partial) negative generator of the time-evolution, is {\em not}
sufficient for the prediction of an instability.
As a counter-example we may consider a 1+1 dimensional black hole with 
$ds^2=h\,dt^2-dr^2/h$.
In this situation there are no angular terms and thus the interior as
well as the exterior $\cal K$-operator are both non-negative.
Consequently the equation of motion is completely stable.
Of course, the interior Hamiltonian $\hat H_<$ displays a global minus
sign, but this does not affect the equation of motion  
\begin{eqnarray}
\hat H
=
\hat H_>+\hat H_<
=
\frac{1}{2}\sumint{\Gamma,>}
\left(
\hat P_{\Gamma,>}^2+
\Omega^2_{\Gamma,>}\hat Q_{\Gamma,>}^2
\right)
-
\frac{1}{2}\sumint{\Gamma,<}
\left(
\hat P_{\Gamma,<}^2+
\Omega^2_{\Gamma,<}\hat Q_{\Gamma,<}^2
\right)
\,.
\end{eqnarray}
Moreover, although the total Hamiltonian is unbounded from above and
below, it splits up into two independent parts which are bounded.
The existence of a horizon is essential for this bifurcation.
In a flat space-time the Wightman \cite{wightman} axioms 
(spectral condition) demand a non-negative generator for stability.
  
The Schwarzschild metric $ds^2=h\,dt^2-dr^2/h$
or $ds^2=h\,dt^2-dr^2/h-r^2\,d\Omega^2$
possesses a unique analytic continuation to values
of $r$ beyond the horizon $r<R$.
In contrast the analytic continuation of the Rindler metric to
negative values of $\rho$  does not lead to a degenerated signature and
complex values of $\rho$ and/or $\tau$ do not describe a physical sheet of the
space-time.
As a consequence one observes {\em no} instability in the Rindler metric
-- i.e. the scenario of the Unruh effect.

The notion of the unstable behaviour obtained above
refers to the time $t$ measured by an (outside) observer at a 
fixed spatial distance to the black hole.   
One might argue that this time coordinate is not capable for describing 
effects inside the black hole due to the coordinate singularity at $r=R$.
However, the instability obtained above is not restricted to the Schwarzschild
coordinates -- it occurs in other coordinate systems as well:
By virtue of the transformation 
\begin{eqnarray}
\label{PGL-S}
dt\,\rightarrow\,dt\pm\frac{\sqrt{R/r}}{1-R/r}\,dr
\,.
\end{eqnarray}
the metric of the black hole can be cast into the
Painlev{\'e}-Gullstrand-Lema{\^\i}tre \cite{Pa,Gu,Le} form
\begin{eqnarray}
\label{PaGuLe}
ds^2=\left(1-\frac{R}{r}\right)dt^2\pm2\sqrt{\frac{R}{r}}\,dt\,dr-
dr^2-r^2\,d\Omega^2
\,.
\end{eqnarray}
This metric is regular everywhere except at the singularity at $r=0$.
The transformation of the unstable solutions in Eq. (\ref{unstable}) 
into this coordinate system via Eq. (\ref{PGL-S}), 
i.e. $t \rightarrow t\pm\Xi(r)$, 
merely results in a simple $r$-dependent factor
\begin{eqnarray}
F_\Gamma(\f{r},t)=
\exp\left(\sqrt{|\lambda_{\Gamma}|}\,t\right)\,f_\Gamma(\f{r})\,
\exp\left(\pm\sqrt{|\lambda_{\Gamma}|}\,\Xi(r)\right)
\,,
\end{eqnarray}
while the unstable behaviour persists.
The same holds true for the Eddington-Finkelstein coordinates with $v=t+r_*$
\begin{eqnarray}
\label{eddi}
ds^2=\left(1-\frac{R}{r}\right)dv^2-2dvdr-r^2d\Omega^2\,.
\end{eqnarray}
Within these coordinates ingoing light rays are simply governed by 
$v=\rm const$.
Both coordinate systems lead to a stationary (but not static) metric,
i.e. the evolution parameter still coincides with a Killing vector.
(This is not the case for the Kruskal metric.)
In summary the instability of the field equation inside the black hole
turns out to be a quite general phenomenon.

\subsection{Back-reaction}\label{back}

The Eddington-Finkelstein metric in Eq. (\ref{eddi}) allows for a
demonstrative visualisation of the unstable behaviour: If one emits
radially ingoing light pulses in uniform intervals these beams are
labelled by equidistant values of $v$. According to the results of the
previous Section the amplitude of the field $\Phi$ inside the black
hole increases exponentially with rising numbers $v$ of the light
rays. Hence we may draw the conclusion that the instability is not
just an artifact caused by an inappropriate description but a physical
effect.      

Nevertheless, for an eternal black hole the outside observer is
completely causally separated from the region of the instability.
Hence the interpretation of the unstable behaviour is not obvious in
that case. But if one considers the possibility of the decay of the
black hole (no matter whether via evaporation or explosion) 
and assumes that this decay can be described using one of the
coordinates above the unstable behaviour should be relevant.
(Of course, the assumption of an eternal black hole automatically 
excludes some of the scenarios where the instability may become
relevant.) 

In order to investigate the consequences of the instability one has to 
deal with the back-reaction problem.
Within all of our previous considerations the quantum field was
regarded as a test field, i.e. it did not influence the given
(externally prescribed) space-time. 
It is known from classical field theory (see e.g. \cite{burko} and
references therein) that the formation of the horizon and the
singularity may well be affected by the scalar field $\Phi$.
(For quantum field theory one expects that the back-reaction will
become important at the Planck scale.)
However, Ref. \cite{burko} deals with radially symmetric fields only. 
For that reason the unstable behaviour was not obtained there. 
The correct implementation of the back-reaction of a quantum field
has to be determined by an underlying theory unifying gravitational
and quantum effects. Since we have no well-established solution to
this problem, we may only speculate about the impact of the quantum
field on the metric based on physical reasonable arguments.           
There are several possible consequences:

\begin{itemize}

\item The explosion of the complete black hole

The unstable field modes evolve as
$\exp(\sqrt{|\lambda_{\Gamma}|}[t-r_*])$.
Hence they "reach" after a finite period of time the Planck scale
vicinity of the horizon, where the classical treatment of the gravitation
is expected to break down. 
In that case one might imagine that the "wave front" destroys the
horizon  and thus the complete black hole. 
(Such an event might perhaps be regarded as a toy candidate for the
big bang.)
In view of arguments concerning the time-reversal symmetry there is no 
obvious reason why the explosion of the complete black hole should be
impossible.  

As long as there is some matter falling into the black hole its
horizon increases. Depending on the particular dynamics of the metric
this may prevent the "wave front" from "reaching" the vicinity of the
horizon. But for a static black hole there is no way to avert the
impact. 

One should be aware that most of the theorems of classical general
relativity -- e.g. the black hole analogues of the laws of
thermodynamics -- are based on appropriate energy conditions,
cf. \cite{ellis}.   
But incorporating the expectation value of the energy-momentum tensor of
the quantum field these energy conditions do not hold in general.
In some cases one may employ averaged energy conditions instead,
but even the validity of an averaged condition is by no means obvious
in view of the unstable solutions of the field equation.  

\item The prevention of the singularity at $r=0$

One might expect that the impact of the instability is at the origin
$r=0$ much stronger than at the horizon $r=R$.
In fact, also those theorems of general relativity that predict a
space-time singularity after a gravitational collapse are based on
energy conditions.    
Accordingly, taking the back-reaction of the quantum field into
account, the formation of the singularity may perhaps be avoided.   
Instead one might imagine some kind of quasi-oscillations:
Impelled by the (exponentially large) amplitudes of the quantum field,
the matter around the origin blows up, absorbs the excitations of the
field, collapses (while the field repeatedly evolves exponentially), 
and eventually blows up again. 

\item The field does not affect the metric

This possibility cannot be excluded within the framework of quantum
field theory in given (external) space-times.
However, the situation of a completely static black hole 
(neglecting the Hawking effect, which is very small for macroscopic
black holes) seems to be rather strange.
In that case the amplitude of the field exceeds the Planck scale after 
a finite period of time (measured by an outside observer).
Hence one would expect drastic modifications of the space-time. 

\end{itemize}

\subsection{Sonic analogue of black holes}\label{sonic}

In 1980 Unruh \cite{acoustic} discovered a very interesting model for
the kinematics of fields in curved space-times. 
He considered the propagation of sound waves in flowing fluids where
the effective equation of motion assumes the same form as the
Klein-Fock-Gordon equation in curved space-times.    
The effective metric depends on the particular flow profile.
Many investigations have been devoted to this topic during the last years,
see e.g. \cite{visser+acoustic}, the recent work \cite{liberati},
and references therein.

Before discussing the consequences of the results of the previous
section within this scenario we shall repeat the basic ideas: 
The flow of a fluid can be described by its local velocity field
$\f{v}$, its density $\varrho$, and the pressure $p$.
The dynamics of the fluid is governed by the non-linear Euler
equation  
\begin{eqnarray}
\dot{\f{v}}+(\f{v}\nabla)\f{v}+\frac{\nabla p}{\varrho}=\f{f}_{\rm ext}
\,,
\end{eqnarray}
if we neglect the viscosity, and the equation of continuity
\begin{eqnarray}
\dot\varrho+\nabla(\varrho\,\f{v})=0
\,.
\end{eqnarray}
For reasons of simplicity we restrict our further considerations to 
a constant speed of sound $c_{\rm s}$.
This implies the very simple relation between the density and the
pressure $p=c^2_{\rm s}\varrho$.
If we assume an irrotational flow $\nabla\times\f{v}=0$, we may
introduce a generating scalar field $\f{v}=\nabla\Phi$.
Now we linearise the non-linear system of the two equations above
around a fixed background solution via 
\begin{eqnarray}
\Phi & = & \Phi_0+\varepsilon\Phi_1+{\cal O}(\varepsilon^2)\,, \nonumber\\
\f{v} & = & \f{v}_0+\varepsilon\f{v}_1+{\cal O}(\varepsilon^2)\,, \nonumber\\
p & = & p_0+\varepsilon p_1+{\cal O}(\varepsilon^2)\,, \nonumber\\
\varrho & = & \varrho_0+\varepsilon\varrho_1+{\cal O}(\varepsilon^2)\,.
\end{eqnarray}
This enables us to consider the propagation of small perturbations
-- i.e. sound waves -- within a given flow profile.
It turns out \cite{acoustic} that the potential $\Phi_1$ of the fluctuations 
satisfies the Klein-Fock-Gordon equation with the effective (acoustic)
metric 
\begin{eqnarray}
g_{\mu\nu}=\frac{\varrho_0}{c_{\rm s}}
\left(
\begin{array}{cr}
c^2_{\rm s}-\f{v}_0^2 & \f{v}_0 \\
\f{v}_0 & -{\bf 1}
\end{array}
\right)
\,.
\end{eqnarray}
Ergo sound waves in flowing fluids share a lot of interesting features 
with fields in curved space-times.
E.g., the surface of transition from subsonic to supersonic flow
represents the acoustic analogue of a horizon. 
For a stationary and radially symmetric flow this surface possesses
even the properties of an event and an apparent horizon.
(Unfortunately this scenario exhibits the problem of fluid conservation 
at $r=0$ which has to be evaded in some way.)

Selecting a particular velocity profile $\f{v}=\pm\f{r}\sqrt{R/r^3}$
it is possible
\cite{visser+acoustic} to simulate a space-time which obeys 
-- up to a conformal factor $r^{-3/2}$ -- the 
Painlev{\'e}-Gullstrand-Lema{\^\i}tre \cite{Pa,Gu,Le} metric 
in Eq. (\ref{PaGuLe}).
According to the results of the previous section the Klein-Fock-Gordon 
equation possesses unstable solutions inside the black hole.
Consequently, also the sound waves within the supersonic region obey
an instability. 
The conformal factor mentioned above and the coordinate transformation 
in Eq. (\ref{PGL-S}) do not alter this conclusion -- see the remarks in
the previous Section.

In contrast to the "real" black hole, where the consequences of the
instability are not a priori clear (back-reaction problem),
there is no possibility to avoid the instability for the acoustic
black hole models since in that case $t$ denotes the appropriate time
also for an inside observer and the sound waves affect the fluid directly.  

In the theory of fluid dynamics, such an instability is a well-known
indicator for the breakdown of the laminar (irrotational) flow,
see e.g. \cite{landau}.
I.e., that flow does not represent a stable fixed point of the
non-linear equation of motion. Accordingly, any small disturbance
will grow up exponentially until the non-linear regime has been
reached.
(It should be mentioned here that the unstable behaviour obtained above 
is not a downstream instability, cf. \cite{landau}, since the perturbation
increases exponentially also at a fixed radius $r$.) 
In order to investigate the behaviour of the flow after leaving the
unstable fixed point -- e.g. pattern formation or turbulence --
one has to consider the non-linear region.
For quantum fields in curved space-times one expects to reach the
non-linear regime at the Planck scale where the back-reaction 
strongly contributes.

Recalling the outcome of the previous Section the unstable behaviour
of the equation of motion results from the angular derivatives of the
$\cal K$-operator. 
Ergo we may draw the conclusion that the quantum field inside the
black hole as well as the supersonically flowing 
fluid favour a spontaneous breaking of the radial symmetry, similar to the
formation of a vortex in the drain of a basin.

\section{Conclusions}\label{conclusions}

\subsection{Summary}\label{summary}

For a minimally coupled, massless and neutral scalar quantum field $\hat\Phi$ 
propagating in an arbitrary physical complete and causal space-time $M$ that 
possesses a static metric of non-degenerated signature it is possible to 
perform a particle definition via diagonalisation of the Hamiltonian.

Application of this method to the 3+1 dimensional Minkowski space-time yields 
the nonexistence of radial ingoing and outgoing particles.
For the 1+1 dimensional Rindler metric we exactly recover the well-known 
Unruh effect.

If we employ the same formalism in order to investigate a black hole the 
associated space-time splits up into two independent domains, inside and 
outside the horizon, respectively.
Within the presented approach a particle definition can be accomplished 
for the exterior region only. 

The quantum field inside the black hole possesses a highly unstable behaviour. 
The corresponding Hamiltonian is unbounded from above and below.
Accordingly, it is not possible to define a vacuum as its ground state and 
particles as excitations over this state.

This instability is not a remnant of an inappropriate description but
a physical effect. Due to our lack of understanding the unification of
quantum theory and gravity the consequences of this effect are not
altogether clear (back-reaction problem). In view of the sonic
analogues of black holes -- where the unstable solutions go along with
the breakdown of the laminar flow -- one might expect that the
instability indicates (at least) the breakdown of the treatment of
quantum fields in given (externally prescribed) space-times.  

\subsection{Discussion}\label{disc}

In order to elucidate the outcome of the formalism presented in this article 
it might be interesting to discuss the main statements together with their 
relations to other approaches: 

As we have observed in Section \ref{ingoing}, the particle definition via 
diagonalisation of the Hamiltonian (equivalent to the energy) does not allow
for the introduction of ingoing and/or outgoing particles in the Minkowski
space-time. The same holds true for more general regular space-times.
As a consequence, the vacuum coinciding with the ground state cannot be 
defined as that state that is annihilated by the "operators" $\hat A_\Gamma$
corresponding to purely ingoing (and/or outgoing) components
$\exp(-i \omega v)/r$ (and/or $\exp(-i \omega u)/r$), with 
$v=t+r$ and $u=t-r$.
Instead the ground state gets annihilated by "operators"
(strictly speaking, operator-valued distributions) corresponding to 
standing waves, i.e. superpositions of ingoing and outgoing components
with equal weights.  
Ergo, considering the collapse of a star to a black hole the initial 
ground state cannot be uniquely and consistently defined by the requirement 
"no ingoing/incoming particles/radiation". 
Ref. \cite{unruh} states explicitly:
{\em Note that we have not defined the vacuum by minimizing
some positive-definite-operator expectation value
(e.g. the Hamiltonian), but we have defined the vacuum as the state
with no incoming particles.}
In order to investigate the relationship of the state defined in this way 
and the initial ground state additional considerations are necessary. 

In contrast to the Minkowski case the ground state of the quantum field 
in the exterior black hole space-time -- the Boulware state -- 
has to be defined via demanding that the action of the annihilators for 
{\em both}, the ingoing {\em and} outgoing modes, yields zero:
$\forall_{\xi\omega\ell m}\,:\,\hat A_{\xi\omega\ell m}\ket{\Psi_{\rm B}}=0$.
This fact illustrates the bifurcation caused by the formation of the horizon.

However, if we assume the black hole to be enclosed by a large 
sphere with e.g. Dirichlet boundary conditions then the definition of
ingoing or outgoing particles is impossible again.
This observation demonstrates manifestly that the particle interpretation 
is a global concept -- it may be influenced by objects (e.g. the sphere) 
at arbitrarily large distances.
 
As another difference between the black hole and the Minkowski situation
we may recall the fact that the $\cal K$-operator of the black hole possesses
-- even in the presence of a finite sphere -- a continuous spectrum.
Due to the additional effective infinity at the horizon the infinite volume
divergence of the black hole space-time {\em cannot} be regularised by 
enclosing it by a finite box. 
(This regularisation applies only to space-time without any horizon.)

There are two main interpretations of the Hawking effect:
The first view considers the particles to be produced by the
dynamics of the space-time during the collapse while within the second
view the radiation is created in a steady rate after the
collapse.
The observations in Section \ref{schwarz}, i.e. the splitting of the
total Hamiltonian into two independent parts and the diagonalisation
of the exterior part by a suitable particle definition
(where the number of particles is conserved), supports the
former interpretation.   

The Hawking effect may be regarded as the verification of the extension of
the laws of thermodynamics to objects like black holes.
This effect allows us to assign a temperature to the black hole via
$T=1/(8M)$ for the Schwarzschild black hole with $h=1-2M/r$.
As a result the associated heat capacity of the black hole turns out
to be negative: If the mass/energy increases the temperature decreases.
The classical laws of thermodynamics predict that an object obeying 
a negative heat capacity will be unstable.
Accordingly, the instability of the black hole interior
as observed in Section \ref{int} might also be regarded as
a verification of the application of thermodynamics to black holes. 

The consequences of the unstable behaviour of the Klein-Fock-Gordon equation
in the interior of the black hole  cannot be deduced rigorously within the 
framework of quantum fields in (externally prescribed) space-times.
The evaluation of the impact of this instability demands the knowledge of the
back-reaction which has to be determined by a unifying theory.
Nevertheless, if the underlying theory possesses an evolution parameter
corresponding to the Schwarzschild time $t$ 
(or one of the other coordinates discussed in Sec. \ref{int})
and contains the treatment of
quantum fields and external metrics in some limiting case, then   
one would expect that the representation of a black hole also obeys the
linearly unstable behaviour.
(This would be consistent with the frequently adopted interpretation 
that black holes are highly excited states of the unifying theory.)

For the situation of the acoustic black hole the interpretation of the 
unstable behaviour is more obvious.
Without any mechanism preserving (enforcing) the radial symmetry 
(e.g. effects of super-fluids) it is probably impossible to realise
the sonic analogue of a black hole experimentally.

\subsection{Outline}\label{out}

The particle definition presented in this article is restricted to
static  space-times. 
This includes the Schwarzschild  and the Reissner metric, but not the Kerr
space-time describing a rotating black hole.
Accordingly, further investigations should be devoted to the extension of the 
previous results to stationary metrics.
(Without any Killing vector generating the time-translation symmetry it is 
probably impossible to perform a unique and physical reasonable particle 
definition.)

Another important extension of the provided formalism
is given by the incorporation of the electromagnetic field
\begin{eqnarray}
{\cal L}=\frac{1}{4}\,F_{\mu\nu}F^{\nu\mu}
\,.
\end{eqnarray}
The Maxwell theory possesses primary and secondary constraints 
\cite{teitelboim}.
These gauge-problems have to be solved before the quantisation
and the particle definition becomes possible.
One way to accomplish this 
-- which seems to be suitable to the canonical approach -- 
is the method of separation of variables \cite{quant-rad}.
Nevertheless there is no obvious reason why the main results of this article 
should not persist.
The equation of motion of the electromagnetic field is given by
\begin{eqnarray}
\nabla_\mu\,F^{\mu\nu}
=
\frac{1}{\sqrt{-g}}\,\partial_\mu\left(
\sqrt{-g}\,g^{\mu\rho}\,g^{\nu\sigma}\,\partial_{\rho}\,A_{\sigma}
\right)
-
\frac{1}{\sqrt{-g}}\,\partial_\mu\left(
\sqrt{-g}\,g^{\mu\rho}\,g^{\nu\sigma}\,\partial_{\sigma}\,A_{\rho}
\right)
=0
\,.
\end{eqnarray}
For a very rough estimate one may drop the second term, 
which is related to the longitudinal degrees of freedom.
The remaining equation possesses unstable interior solutions similar
to the scalar field scenario.   

The investigation of the Dirac field  
\begin{eqnarray}
{\cal L}=\overline{\Psi}\left(\frac{i}{2}\,
\gamma^\mu\stackrel{\leftrightarrow}{D_\mu}
-m\right)\Psi
\end{eqnarray}
around charged black holes creates some new kind of problems,
see e.g. \cite{khriplovich}.
Similar to the Schwinger mechanism in the semi-classical description a 
tunnelling process is possible.
This tunnelling probability gives raise to the question of whether
a stable vacuum in the quantum field theoretical treatment exists. 

Having obtained a linear instability of the linear equations of motion 
one may ask whether the unstable behaviour persists
for non-linear equations of motion including interaction terms, for
instance $\Phi^4$.
One might suspect that the non-linear terms generate new stable fixed
points of the equation of motion -- i.e. a non-perturbative
stabilisation of the black hole.
However, in this case the amplitude of the field has to be located at
some fixed scale while the (linear) instability exists for arbitrary
large scales $|\lambda_\Gamma|$.  
This might be an argument for the dominance of the unstable linear
contribution in this region.
In order to elucidate this point it is necessary to consider the scale 
behaviour of the interacting theory.

This article considers the propagation of quantised fields in a given 
(i.e. externally prescribed) space-time. 
To examine how the quantum fields influence the metric one has to deal 
with the back-reaction problem.
Within the canonical (operator) quantisation one usually employs the 
renormalised expectation value of the energy-momentum tensor as the 
source of the Einstein equations \cite{wald-t}, and within the path-integral
approach one may integrate out the quantum field in order to obtain an 
effective action (accounting for the degrees of freedom associated
with the dynamics of the space-time). 
However, a complete solution to this question probably requires the 
knowledge of the unification of general relativity and quantum field theory.  

\newpage

\section*{Acknowledgement}

The author is indebted to 
A. Calogeracos,
K. Fredenhagen,
I. B. Khriplovich, 
G. Plunien, 
G. Soff,
and R. Verch 
for fruitful conversations and helpful criticism.
Discussions with R. Picard concerning questions of
functional analysis are also gratefully acknowledged.
This work was partially supported by BMBF, DFG and GSI.



\end{document}